\documentclass[aps,prb,twocolumn]{revtex4-2}
\usepackage{amssymb}
\usepackage{graphicx}
\usepackage{epsf,epsfig}
\usepackage{bm}
\usepackage{bbm}
\usepackage{amsfonts}
\usepackage{amsmath}
\usepackage{graphics}
\usepackage[normalem]{ulem}
\usepackage{xcolor}
\usepackage{hyperref}
\usepackage{cancel}

\newcommand{\be}{\begin{eqnarray}}
\newcommand{\ee}{\end{eqnarray}}
\newcommand{\ba}{\begin{array}}
\newcommand{\ea}{\end{array}}
\newcommand{\no}{\nonumber}
\newcommand{\tr}{\mbox{tr}}
\newcommand{\Tr}{\mbox{Tr}}

\newcommand{\eps}{\varepsilon}

\newcommand{\bfr}{{\bf r}}

\newcommand{\bfp}{{\bf p}}
\newcommand{\bfq}{{\bf q}}
\newcommand{\bfk}{{\bf k}}

\newcommand{\op}{\operatorname}
\newcommand{\omg}{\omega}
\newcommand{\mD}{\mathcal{D}}
\newcommand{\mI}{\mathcal{I}}
\newcommand{\ul}{\underline}

\begin{document}

\title{Low-temperature thermoelectric transport in the disordered two-dimensional electron gas}

\author{Zahidul Islam Jitu}
\author{Georg Schwiete}

\affiliation{Department of Physics and Astronomy, The University of Alabama, Tuscaloosa, Alabama 35487, USA}

\begin{abstract}
We study interaction corrections to the thermoelectric transport coefficient $\alpha$ and the thermopower $S$ in the two-dimensional disordered electron gas with long-range Coulomb interactions. To this end, we analyze the heat density-density correlation function using the recently derived generalized non-linear sigma model with particle-hole asymmetry. We present a detailed account of the perturbative calculation of the correlation function, the analysis of its structure, and the derivation of the final results for the transport coefficients. We show that both real and virtual processes give rise to logarithmic interaction corrections to $\alpha$ and $S$ with $\delta \alpha/\alpha<0$ and $\delta S/S<0$ in the diffusive transport regime $T\ll 1/\tau$, where $T$ is the temperature and $1/\tau$ is the disorder scattering rate. A short account of our findings is presented in an accompanying letter.
\end{abstract}

\maketitle

\setcounter{secnumdepth}{4}

\section{Introduction}

The thermoelectric transport coefficient $\alpha$ characterizes the ability of a system to conduct an electric current ${\bf j}$ in response to a temperature gradient, ${\bf j}=-\alpha\nabla T$ \cite{a_RemarkOnsager}. When both a temperature gradient and an electric field ${\bf E}$ are simultaneously present, the linear response relation for the electric current reads as ${\bf j}=\sigma {\bf E}-\alpha\nabla T$, where $\sigma$ is the electric conductivity. Instead of measuring the thermoelectric transport coefficient directly, it is experimentally more convenient to find the thermopower $S=\alpha/\sigma$, also known as the Seebeck coefficient. The thermopower can be obtained as $S=|{\bf E}|/|{\nabla T}|$ in an experimental set up with vanishing electric current, ${\bf j}=0$.

This manuscript is concerned with the calculation of quantum corrections to $\alpha$ and $S$ originating from the interplay of disorder and interactions for the disordered two-dimensional electron gas. Corrections of this type arise for various thermodynamic quantities and transport coefficients and are known as Altshuler-Aronov corrections or interaction corrections \cite{Altshuler80,Finkelstein83,Altshuler85}. Here, we are interested in the low-temperature diffusive transport regime $T\ll 1/\tau$ \cite{notation_remark}, where $1/\tau$ is the scattering rate. We only treat the purely electronic contribution to thermopower, the so-called diffusion thermopower. The diffusion thermopower is expected to prevail over to the phonon drag contribution at low temperatures. Our findings have already been summarized in a compact form in Ref.~\cite{jitu24}, hereafter referred to as paper I. In this manuscript, we provide details of a perturbative analysis of interaction corrections to the heat density-density correlation function for the disordered two-dimensional electron gas with long-range Coulomb interactions. Knowledge of this correlation function allows us to determine the thermoelectric transport coefficient and the thermopower. A similar analysis has been performed for the calculation of the thermal conductivity from the heat density-heat density correlation function in Ref.~\onlinecite{Schwiete16a}. We devote particular attention to logarithmic corrections from the sub-thermal energy interval $(T^2/D\kappa_s^2,T)$, where $D$ is the diffusion coefficient and $\kappa_s$ is the inverse screening radius. If present, these corrections can compete with the logarithmic corrections from the renormalization group (RG) energy interval $(T,1/\tau)$. It is well known that processes with energies below the temperature do not give rise to interaction corrections to the electric conductivity \cite{Altshuler80,Finkelstein83,Altshuler85}. Such processes are, however, responsible for Wiedemann Franz law-violating corrections to the thermal conductivity in the presence of long-range Coulomb interactions \cite{Livanov91,Raimondi04,Niven05,Catelani05,Michaeli09,Schwiete16a,Schwiete16b}. In the previous calculation of the heat density-density correlation function in Ref.~\onlinecite{Fabrizio91}, contributions from the sub-thermal energy interval were not taken into account. Here, we demonstrate how the logarithmic corrections from the RG and the sub-thermal energy intervals are compatible with the general form of the heat density-density correlation function and contribute to the overall corrections to $\alpha$ and $S$.

As discussed in detail in Ref.~\onlinecite{Schwiete21} and paper I, the calculation of $\alpha$ and $S$ needs to be performed with a higher accuracy compared to $\sigma$ and $\kappa$ due to the importance of particle-hole asymmetry. This is already obvious from the Drude-Boltzmann result $\alpha=(2\pi^2/3)eT(\nu_\eps D_\eps)'$, where $\nu_\eps$ and $D_\eps$ denote the frequency-dependent density of states and diffusion coefficient, respectively, the prime denotes the derivative with respect to the frequency $\eps$ and and $e$ is the electron charge. The relevance of particle-hole asymmetry manifests itself through the sensitivity of $\alpha$ to the energy-dependence of quantities such as the velocity $v_\eps$ and scattering time $\tau_\eps$ entering the diffusion coefficient $D_\eps=v_\eps^2\tau_\eps/2$, or the density of states. In two dimensions ($2d$), and for a quadratic dispersion that we study here, the density of states and the scattering time are approximately constant \cite{Fabrizio91,Schwiete21}, and it is sufficient to include the frequency-dependence of the velocity encoded in the diffusion coefficient. Recently, a generalized Finkel'stein non-linear sigma model (NL$\sigma$M) \cite{Finkelstein90} with particle-hole asymmetry has been derived for this system \cite{Schwiete21}. This NL$\sigma$M explicitly incorporates the particle-hole asymmetry through the parameter $D'_\eps$ and forms the basis for our considerations. 

This manuscript is structured as follows. In Sec.~\ref{sec:chikngeneral}, we introduce the heat density-density correlation function and discuss the connection between this correlation function and the thermoelectric transport coefficient $\alpha$. We also perform a perturbative expansion that will later allow us to identify interaction corrections to the transport coefficients. In Sec.~\ref{sec:Calculation}, we turn to the perturbative calculation of the correlation function. Specifically, in Sections~\ref{sec:basic} and ~\ref{sec:param}, we introduce the NL$\sigma$M formalism in the presence of particle-hole asymmetry. Subsequently, we present a detailed account of the individual corrections to the correlation function and the thermoelectric transport coefficient. In Sec.~\ref{sec:thermo}, we obtain the final expressions for the corrections to the thermopower $S$ and discuss how interaction corrections affect the thermoelectric figure of merit. We conclude the manuscript in Sec.~\ref{sec:conclusion}. Additional technical details of the calculation are included in Appendices~\ref{app:gen}-\ref{sec:Jterms}.

\section{The heat density-density correlation function in the diffusive regime}
\label{sec:chikngeneral}

Our calculation of the thermoelectric transport coefficient $\alpha$ is based on the heat density-density correlation function. There are two main reasons for this choice. First, the form of this correlation function is highly constrained. The constraints originate from the conservation laws for the particle number and energy, and from the relation of the static limit of the correlation function to the thermodynamic susceptibility $\zeta=\partial_T n$. These constraints provide very valuable checks for the calculation of the interaction corrections. Second, certain contributions to the heat density-density correlation function bear a close similarity with the heat density-heat density correlation function that has been used for the calculation of the thermal conductivity in Ref.~\onlinecite{Schwiete16a}. We now proceed to discuss the general structure of the correlation function and its relation to the thermoelectric transport coefficient.

\subsection{General structure}
The heat density-density correlation function in diffusive regime assumes the following form \cite{Fabrizio91}
\begin{align}
\chi_{kn}(\bfq,\omega)=& \frac{\chi_{kn}^{st}D_nD_k\bfq^4+i\omega\bfq^2L}{(D_n\bfq^2-i\omega)(D_k\bfq^2-i\omega)}.\label{eq:chikndiff}
\end{align}
A derivation of this result is provided in Appendix~\ref{app:gen}. The coefficient $L$ can be found from the correlation function as $L = - \lim_{\omega \to 0} \lim_{\bfq \to 0} \frac{\omg}{\bfq^2} \Im\left[\chi_{kn}(\bfq,\omg)\right]$ \cite{Fabrizio91}. The thermoelectric transport coefficient $\alpha$ is related to the parameter $L$ appearing in Eq.~\eqref{eq:chikndiff} as $L=T\alpha/e$ (after reintroducing the charge $e$ omitted in the derivation presented in Appendix~\ref{app:gen}). Further, $D_n$ and $D_k$ are the charge and heat diffusion coefficients, respectively, and $\chi_{kn}^{st}$ is the static part of the correlation function.

The form of the heat density-density correlation function presented in Eq.~\eqref{eq:chikndiff} is consistent with the following two constraints,
\begin{align}
\chi_{kn}(\bfq=0,\omega \rightarrow 0)=& 0,
\\
\chi_{kn}(\bfq \rightarrow 0,\omega = 0)=& \chi_{kn}^{st}.
\end{align}
Here, the first relation is a consequence of the particle number and energy conservation. As for the second relation, it is worth mentioning that the static part of the correlation function is related to the thermodynamic susceptibility $\zeta=\partial_Tn$ as $\chi_{kn}^{st}=-T\zeta$, see Appendix~\ref{app:gen}.

\subsection{Perturbative expansion}
\label{sec:struc}

As can be seen from Eq.~\eqref{eq:chikndiff}, the diffusive form of the heat density-density correlation function is characterized by four distinct parameters, $\chi_{kn}^{st}$, $D_n$, $D_k$ and $L$. At low temperatures, these parameters acquire interaction corrections, which we denote as $\delta \chi_{kn}^{st}$, $\delta D_n$, $\delta D_k$ and $\delta L$. In this work, we are interested in a parameter regime for which the interaction corrections are small and a first order expansion of the correlation function is justified 
\begin{align}
\chi_{kn}(\bfq,\omega) &\approx \chi^{(0)}_{kn}(\bfq,\omega) +\delta \chi_{kn}^{st} \label{eq:1st}\\
&+ i \omg \mD^{(n)}_{\bfq,\omg} \delta \chi_{kn}^{st}\no \\
&+ i \omg \mD^{(k)}_{\bfq,\omg} \mD^{(n)}_{\bfq,\omg} \bfq^2 D_n \delta \chi_{kn}^{st} \no
\\
&+ i \omg \mD^{(n)}_{\bfq,\omg} \mD^{(k)}_{\bfq,\omg} \bfq^2 \delta L \no\\
&- i \omg (\mD^{(n)}_{\bfq,\omg})^2 \mD^{(k)}_{\bfq,\omg} \bfq^4 \delta D_n L\no
\\
&- i \omg \mD^{(n)}_{\bfq,\omg} (\mD^{(k)}_{\bfq,\omg})^2 \bfq^4 \delta D_k L. \no
\end{align}
Here, $\mD^{(n)}_{\bfq,\omg} = (D_{n}\bfq^2-i\omega)^{-1}$ and $\mD^{(k)}_{\bfq,\omg} = (D_{k}\bfq^2-i\omega)^{-1}$ are charge and heat diffusion modes, respectively. 

For the specific case of a two-dimensional disordered electron gas with a quadratic dispersion, the following initial parameter values can be obtained from conventional {Drude-Boltzmann theory: $D_n = D_k = D$, $\chi^{st}_{kn} = 0$, $\delta \chi^{st}_{kn} = - c_0 T \partial_\mu z$, and $L = c_{0} T D_\eps'$, where $D = v_F^2\tau/2$ is the diffusion coefficient, $D_\eps' = {D}/{\mu}$,} and $c_0 = 2 \pi^2 \nu_0 T/3$ represents the specific heat. As  discussed in paper I, the interaction corrections to the heat and charge diffusion coefficients can be presented as $\delta D_n = \delta D$ and $\delta D_k = \delta D + \frac{1}{2} D I^h - D \delta z$ \cite{Finkelstein83,Castellani84,Castellani87,Schwiete16a}. Here, the correction to the (bare) diffusion coefficient $\delta D$ and the deviation $\delta z=z-1$ of Finkel'stein's frequency renormalization $z$ from $1$ represent interaction corrections from the RG energy interval. The correction to the heat diffusion coefficient $D_k$ also acquires a contribution from the sub-thermal energy interval, which is proportional to the logarithmic integral $I^h$  (see Appendix~\ref{sec:int} for details). Inserting the initial parameter values for the disordered two-dimensional electron gas and the well-known interaction corrections $\delta D_n$ and $\delta D_k$ into Eq.~\eqref{eq:1st}, we arrive at 
\begin{align}
\delta\chi_{kn}(\bfq,\omega) &= \delta \chi_{kn}^{st} + 2 i \omg \mD_{\bfq,\omg} \delta \chi_{kn}^{st} \label{eq:strucPert} \\
&+ i \omg \mD_{\bfq,\omg}^2 \bfq^2 \left(\delta L/L + \delta z - I^h/2\right)L \no\\
&- \omg^2 \mD_{\bfq,\omg}^2 \delta \chi_{kn}^{st} - \omg^2 \mD_{\bfq,\omg}^3 \bfq^2 \left(\delta z - I^h/2\right) L\no \\
& - 2 i \omg \mD_{\bfq,\omg}^3 \bfq^4 \delta D L, \no
\end{align}
where $\mD_{\bfq,\omg} = (D \bfq^2-i \omega)^{-1}$, $\delta\chi_{kn}(\bfq,\omega) = \chi_{kn}(\bfq,\omega) - \chi_{kn}^{(0)}(\bfq,\omega)$, and $\chi_{kn}^{(0)}(\bfq,\omega) = i \omg \bfq^2 L \mD_{\bfq,\omg}^2$. A careful comparison of Eq.~\eqref{eq:strucPert} to the results of the perturbative calculation of $\chi_{kn}$ detailed in Sec.~\ref{sec:Calculation} will allow us to extract $\delta L$. The correction $\delta L$, in turn, is closely related to the interaction corrections to the thermoelectric transport coefficient, $\delta \alpha= e \delta L / T$ and the thermopower $\delta S=e (\delta L \sigma - L \delta \sigma)/(T \sigma^2) $.

Due to the complexity of the calculation, it will sometimes be useful to further classify the corrections according to their frequency and momentum structure. We define
\begin{align}
\delta \chi_{kn}(\bfq,\omega) &= \delta \chi_{kn}^{st} + \delta \chi^A_{kn}(\bfq,\omega) + \delta \chi^B_{kn}(\bfq,\omega) \no\\
&+ \delta \chi^C_{kn}(\bfq,\omega) + \delta \chi^D_{kn}(\bfq,\omega) + \delta \chi^E_{kn}(\bfq,\omega),\label{eq:chidecomp}
\end{align}
with
\begin{align}
\delta \chi^A_{kn}(\bfq,\omega) &= 2 i \omg \mD_{\bfq,\omg} \delta \chi_{kn}^{st},\label{eq:chiAtoE}\\
\delta \chi^B_{kn}(\bfq,\omega) &= i \omg \bfq^2 \mD_{\bfq,\omg}^2 \left(\delta L/L + \delta z - I^h/2\right)L,\no
\\
\delta \chi^C_{kn}(\bfq,\omega) &= - \omg^2 \mD_{\bfq,\omg}^2 \delta \chi_{kn}^{st}, \no\\ 
\delta \chi^D_{kn}(\bfq,\omega) &=- \omg^2 \bfq^2 \mD_{\bfq,\omg}^3 \left(\delta z - I^h/2\right) L, \no
\\
\delta \chi^E_{kn}(\bfq,\omega) &= - 2 i \omg \bfq^4 \mD_{\bfq,\omg}^3 \delta D L. \no
\end{align}
This decomposition of $\delta \chi_{kn}(\bfq,\omega)$ is not unambiguous. For example, the sum of the terms proportional to $I^h$ acquires the form of $\delta \chi^E_{kn}(\bfq,\omega)$ after applying the relation $1+i\omega \mathcal{D}_{\bfq,\omega}=D\bfq^2 \mathcal{D}_{\bfq,\omega}$. As we will see, however, the decomposition introduced in Eq.~\eqref{eq:chiAtoE} lends itself naturally to our explicit calculation of the interaction corrections.

\section{Calculation of the heat density-density correlation function from the generalized NL{\boldmath$\sigma$}M}
\label{sec:Calculation}

\subsection{Basic formalism}
\label{sec:basic}

Our calculation of the heat density-density correlation function is based on the relation
\begin{align}
\chi_{kn}(x_1,x_2) = \frac{i}{2} \frac{\delta^2 \mathcal{Z}}{\delta \eta_2(x_1) \delta \phi_1(x_2)} \bigg|_{\eta=0,\phi=0} \label{chiDer}
\end{align}
where $\mathcal{Z} = \int DQ \exp(i S)$ is the partition function of the generalized Keldysh NL$\sigma$M with particle-hole asymmetry \cite{Schwiete21}. The action $S=S_F+S_M$ incorporates the gravitational potential $\eta$ \cite{Luttinger64} and the scalar potential $\phi$ as source fields and has already been defined in paper I. We repeat the action here for convenience 
\begin{align}
S_F[\hat{X}]=&\frac{i \pi\nu_0 }{4}\Tr\left[ D(\nabla \underline{\hat{X}})^2+4i \hat{\eps}^\eta_{\varphi} \underline{\delta \hat{X}} \right] \label{eq:SF}
\\
&-\frac{\pi^2\nu_0^2}{4}\int_{\bfr,\bfr',t} \tr[\hat{\gamma}_i\hat{\lambda}(\bfr,t) \underline{\delta \hat{X}_{t t}}(\bfr)] \hat{\gamma}_2^{ij}\nonumber
\\
&\quad\quad\quad\times V_s(\bfr-\bfr') \tr[\hat{\gamma}_j \underline{\delta \hat{X}_{t t}}(\bfr')],\nonumber
\\
S_M[\hat{Q}]=&\frac{\pi\nu_0}{16}DD_\eps'\Tr[\nabla^2 \hat{Q}(\nabla \hat{Q})^2]. \label{eq:SM}
\end{align}
with $\hat{X}=\hat{Q}+\frac{1}{4i}D_\varepsilon'(\nabla \hat{Q})^2$. We refer the reader to paper I for further details regarding the action and its individual terms. A peculiarity of the generalized NL$\sigma$M is that the terms with prefactor $D'_\varepsilon$ are formally much smaller than those present in the conventional Finkel'stein model $S_F[\hat{Q}]$. These terms need to be included in the action because their symmetry differs from the conventional terms and a finite thermopower can only be obtained in their presence as discussed in detail in Ref.~\cite{Schwiete21}. It is worth mentioning that similar arguments arise in the context of the scaling theory of the Hall coefficient near the metal-insulator transition discussed in Ref.~\cite{Wang94} as well as a recently introduced NL$\sigma$M approach to disordered systems with intrinsic spin-orbit coupling \cite{Virtanen22}.

Interaction corrections to the static part $\chi^{\op{st}}_{kn}=\chi_{kn}({\bf q}\rightarrow 0,\omega=0)$ of the correlation function have been addressed in previous studies \cite{Fabrizio91,Schwiete21} and it was found that the result can be expressed as $\delta \chi^{st}_{kn} = - c_0 T \partial_\mu z$. Here, $c_0 = 2 \pi^2 T \nu_0 / 3$ is the specific heat, and $z$ is the frequency renormalization familiar from the RG analysis of the conventional Finkel'stein NL$\sigma$M. Here, we focus on the interaction corrections to the dynamical part of the correlation function $\chi_{kn}^{\op{dyn}}(\bfq,\omg)=\chi_{kn}(\bfq,\omg)-\chi_{kn}^{\op{st}}$.

\begin{figure}[h]
\includegraphics[width=7.5cm]{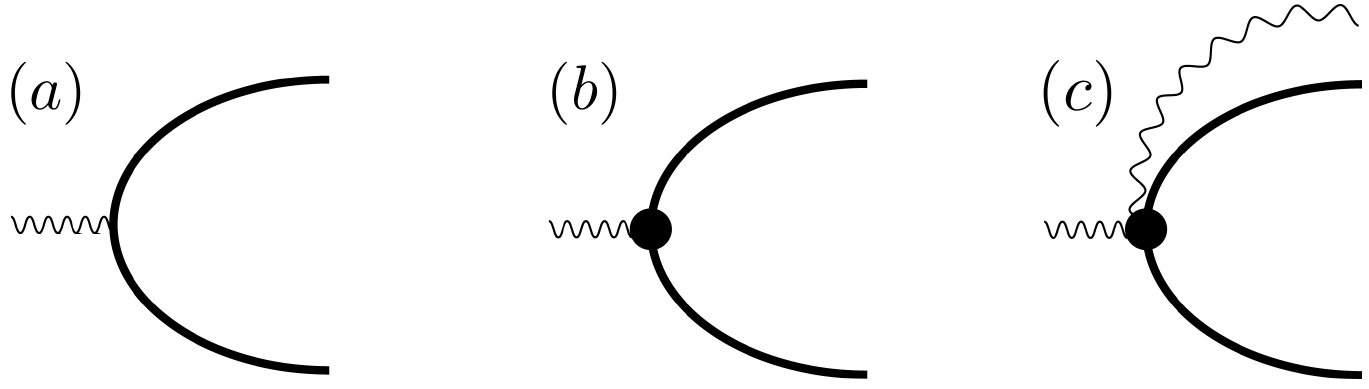}
\caption{Different types of vertices: the density vertex is shown in $(a)$, the frequency vertex in $(b)$, and the interaction vertex in $(c)$.}
\label{fig:freqvert}
\end{figure}

In view of Eq.~\eqref{chiDer}, it is sufficient to expand the action to linear order in the gravitational potential $\hat{\eta}$ (with the scalar source field $\hat{\phi}$ already linear in the action). Employing the expansion $\hat{\lambda} \approx 1 - \hat{\eta}$ in Eqs.~\eqref{eq:SF} and \eqref{eq:SM} leads to the following expression
\begin{align}
&S_{\op{lin}}=\frac{i \pi \nu_0}{4}  \Tr \left[D \left(\nabla \hat{X}\right)^2 \right.\label{eq:Slin} 
\\
&\left. + 2 i \left\{\hat{\eps} - \hat{\phi} , 1- \hat{\eta}\right\} \ul{\delta \hat{X}} +\frac{1}{4 i} D' D \left(\nabla \hat{Q}\right)^2 \nabla^2 Q\right]\no
\\
&-\frac{\pi^2\nu_0^2}{4}\int_{\bfr,\bfr',t} \tr[\hat{\gamma}_i (1-\hat{\eta}(\bfr,t)) \underline{\delta \hat{X}_{t t}}(\bfr)] \hat{\gamma}_2^{ij}\nonumber
\\
&\quad\quad\quad\times V_s(\bfr-\bfr') \tr[\hat{\gamma}_j \underline{\delta \hat{X}_{t t}}(\bfr')].\nonumber
\end{align}
The action in Eq.~\eqref{eq:Slin} contains a term that is quadratic in the source fields ($\eta \phi$). This term is only required for the calculation of the static part of the correlation function and we therefore exclude it from further considerations. The linearized action can then be divided into four parts,
\begin{align}
S_{\op{lin}} &= S_{\hat{\eta} = 0, \hat{\phi}=0} + S_{\hat{\phi}} + S^\eps_{\hat{\eta}} + S^V_{\hat{\eta}}\label{eq:Slin1}
\end{align}
The first term is the sigma model action in the absence of source fields. The remaining three terms are each represented by a different type of vertex. The second term in Eq.~\eqref{eq:Slin1} describes the coupling of the the scalar field to the density, and reads as 
\begin{align}
S_{\hat{\phi}} &= \pi \nu_0 \Tr \left[\hat{\phi} \ul{\delta \hat{X}}\right]. \label{scalarCouple}
\end{align}
The diagrammatic representation of the associated density vertex is displayed in Fig.~\ref{fig:freqvert}(a). When combined, the third and the fourth term in Eq.~\eqref{eq:Slin1} describe the coupling of the gravitational potential to the heat density. 

The third term has an explicit frequency dependence
\begin{align}
S^\eps_{\hat{\eta}} &= \frac{\pi \nu_0}{2} \Tr\left[\left\{\hat{\eps} , \hat{\eta}\right\} \ul{\delta \hat{X}}\right].  \label{freqCouple}
\end{align}
Diagrammatically, this term gives rise to the frequency vertex shown in Fig.~\ref{fig:freqvert}(b). The fourth term in Eq.~\eqref{eq:Slin1} involves the Coulomb interaction
\begin{align}
S^V_{\hat{\eta}} &=\frac{\pi^2\nu_0^2}{4}\int_{\bfr,\bfr',t} \tr[\hat{\gamma}_i \hat{\eta}(\bfr,t) \underline{\delta \hat{X}_{t t}}(\bfr)] \hat{\gamma}_2^{ij}\nonumber
\\
&\quad\quad\quad\times V_s(\bfr-\bfr') \tr[\hat{\gamma}_j \underline{\delta \hat{X}_{t t}}(\bfr')]. \label{intCouple}
\end{align}
The corresponding vertex, which we will refer to as the interaction vertex, is shown in Fig.~\ref{fig:freqvert}(c). All the vertices associated with the heat density, namely the frequency vertex and the interaction vertex, are drawn with a dot, while the density vertices are displayed without the dot in the diagrams. 

Differentiation of the partition function according to Eq.~\eqref{chiDer} generates two separate terms $\chi^{\op{dyn}}_{kn}=\chi^{\op{dyn}}_{\eps n}+\chi^{\op{dyn}}_{Vn}$. The first term involves the frequency vertex and the density vertex and reads as
\begin{align}
\chi^{\op{dyn}}_{\eps n}(x_1,x_2) &= - \frac{i \pi^2 \nu_0^2}{2} \int_{\eps_i}  e^{-i t_1(\eps_1-\eps_2)} \left\langle \bar{\eps}_{12} \right.\no
\\
&\times \left. \tr\left[\hat{\gamma_2} \ul{\delta \hat{X}}_{\eps_1 \eps_2}(\bfr_1)\right] \tr\left[\hat{\gamma_1} \ul{\delta \hat{X}}_{t_2 t_2}(\bfr_2)\right]\right\rangle_{\hat{\eta}=\hat{\phi}=0},\label{eq:chidyneps}
\end{align}
where $\bar{\eps}_{ij} = ({\eps_i+\eps_j})/{2}$ denotes the average frequency. The second term involves the interaction and the density vertices 
\begin{align}
\chi^{\op{dyn}}_{V n}(x_1,x_2) &= - \frac{i \pi^3 \nu_0^3}{8} \int_{\bfr_3} \left\langle \tr \left[\hat{\gamma}_i \ul{\delta \hat{X}}_{t_1 t_1}(\bfr_1)\right] \hat{\gamma}_2^{ij} V_s(\bfr_1-\bfr_3) \right. \no 
\\ 
&\left. \times \tr \left[\hat{\gamma}_j \ul{\delta \hat{X}}_{t_1 t_1}(\bfr_3)\right]
\tr\left[\hat{\gamma_1} \ul{\delta \hat{X}}_{t_2 t_2}(\bfr_2)\right] \right\rangle_{\hat{\eta} = \hat{\phi}=0}. \label{eq:chidynV}
\end{align}
In both of the above formulas, the averaging $\langle\dots\rangle_{\hat{\eta} = 0, \hat{\phi}=0}$ is taken with respect to the action $S_{\hat{\eta} = 0, \hat{\phi}=0}$.

\subsection{Parameterization of $\hat{Q}$ and dynamical screening}
\label{sec:param}
We perform a first-order perturbative expansion of the heat density-density correlation function, for which the dimensionless resistance $\rho=1/(4\pi^2 \nu_0 D)$ serves as the small parameter. For this application, the configurations of the $\hat{Q}$ matrix field with the largest weight consist of small rotations around the metallic saddle point $\hat{\sigma}_3$. These rotations can be parameterized in different ways. Here, we use a continuous family of parameterizations proposed in Ref.~\onlinecite{Ivanov06},
\begin{align}
\hat{Q} = \hat{\sigma}_3 \frac{\left(\frac{\hat{P}}{2} + \sqrt{1+(1-\zeta) \frac{\hat{P}^2}{4}}\right)^2}{1 - \zeta \frac{\hat{P}^2}{4}}, 
\end{align}
with $\{\hat{\sigma}_3, \hat{P}\} = 0$. We will refer to this parameterization as the $\zeta$-parameterization. The $\zeta$-parameterization has a unit Jacobian \cite{Ivanov06}. For our perturbative calculation an expansion up to fourth order is sufficient
\begin{align}
\hat{Q} &\approx \sigma_3 \left(1 + \hat{P} + \frac{\hat{P}^2}{2} + \frac{(1 + \zeta)  \hat{P}^3}{8} + \frac{\zeta \hat{P}^4}{8}\right). \label{parametrizationQ}
\end{align}
The choice to use the $\zeta$-parameterization is motivated by the observation that the parameter $\zeta$ needs to cancel out after adding all terms generated at a given order in the perturbation theory. This cancellation can serve as a useful check for the validity of the calculation. Incorporating the $\zeta$-parameterization also allows us to perform  valuable cross-checks with alternative parameterizations. For example, for $\zeta = 0$ one obtains the so-called square-root-odd parametrization. Setting $\zeta=-1$ generates the so-called square-root even parameterization \cite{hikami1981} for expansions up to fourth order, which aligns well with the requirements of our calculation. This parameterization is known to correspond to the conventional diagrammatic Green's function approach \cite{abrikosov1975}. Setting $\zeta = 1/3$ generates the so-called exponential parametrization \cite{Efetov80} for expansions up to fourth order.

Inserting the parameterization \eqref{parametrizationQ} into the action $S_{\eta=0,\phi=0}$, we immediately obtain the quadratic action $S_0$ of the noninteracting theory
\begin{align}
S_0 = - \frac{i \pi \nu_0}{4} \tr[\hat{D}_{\eps} (\nabla \hat{P})^2 - 2 i \hat{\eps} \hat{\sigma}_3 \hat{P}^2], \label{eq:GaussianAction}
\end{align}
which will serve as the basis of the perturbation theory. Gaussian averages with respect to $S_0$ can easily be performed after presenting the matrix $\hat{P}$ in the following form in Keldysh space \cite{Kamenev11}
\begin{align}
\hat{P}_{\eps \eps'}(\bfr) = \begin{pmatrix}
0 & d^{cl}_{\eps \eps'}(\bfr) \\
d^q_{\eps \eps'}(\bfr) & 0
\end{pmatrix},
\end{align}
where $d^{cl/q}$ are Hermitian matrices. The contraction rule for these matrices reads as
\begin{align}
&\langle d^{cl}_{\alpha \beta;\eps_1 \eps_2}(\bfq) d^{q}_{\gamma \delta;\eps_3 \eps_4}(\bfq') \rangle_0 =\no \\
& -\frac{2}{\pi \nu_0} \mD^\eps_{\bfq,\omega} (2 \pi)^d \delta(\bfq+\bfq') \delta_{\eps_1,\eps_4} \delta_{\eps_2,\eps_3} \delta_{\alpha,\delta} \delta_{\beta,\gamma}
\end{align}
where $\delta_{\eps_1,\eps_4} = 2 \pi \delta(\eps_1-\eps_4)$, $\eps= (\eps_1 + \eps_2)/{2}$, $\omega=\eps_1-\eps_2$, $\alpha,\beta,\gamma,\delta$ are spin indices and the average $\langle\dots \rangle_0$ is respect to the action $S_0$ in Eq.~\eqref{eq:GaussianAction}. The retarded diffuson propagator 
\begin{align}
\mD_{\bfq,\omg}^{\eps} &= \frac{1}{D_\eps \bfq^2 - i \omg} \label{phDiffuson}
\end{align}
accounts for particle-hole asymmetry through the frequency-dependence of the diffusion coefficient, $D_\eps=D (1+\eps/\mu)$, as in Ref.~\cite{Schwiete10,Schwiete13}. The advanced diffusion will be denoted as $\bar{\mD}$, and it is related to the retarded diffuson as $\bar{\mD}_{\bfq,\omg}^{\eps} = \mD_{\bfq,-\omg}^{\eps}$. With this preparation, we can formulate a useful contraction rule for the matrices $\hat{P}$, namely 
\begin{align}
&\langle \tr[\hat{A} \hat{P}_{\eps_1 \eps_2}(\bfr)] \tr[\hat{B} \hat{P}_{\eps_3 \eps_4}(\bfr')]\rangle_0\no \\
& = -\frac{2}{\pi \nu_0} \tr[\hat{A}^\perp \hat{\Pi}_{\eps_1 \eps_2}(\bfr-\bfr') \hat{B}^\perp] \delta_{\eps_1,\eps_4} \delta_{\eps_2,\eps_3}.\label{eq:contraction2}
\end{align}
Here, $\hat{A}^\perp = \frac{1}{2} (\hat{A} - \hat{\sigma}_3 \hat{A} \hat{\sigma}_3)$ is the off-diagonal part of the matrix $\hat{A}$, and
\begin{align}
\hat{\Pi}_{\eps_1 \eps_2}(\bfr-\bfr') = \begin{pmatrix}
\mD^{\eps}_{\omg}(\bfr-\bfr') & 0 \\
0 & \bar{\mD}^{\eps}_{\omg}(\bfr-\bfr')
\end{pmatrix}.
\end{align}

We now turn our attention to the Coulomb interaction. After inserting the parameterization of the $\hat{Q}$-matrix, the part of the action $S$ that contains the Coulomb interaction also generates a quadratic term in $\hat{P}$. The main role of this term is to introduce dynamical screening of the Coulomb interaction into the calculation. In order to account for this effect, we introduce the dynamically screened Coulomb interaction as
\begin{align}
\hat{V}_{\bfk,\nu} = \begin{pmatrix}
V^K_{\bfk,\nu} & V^R_{\bfk,\nu} \\
V^A_{\bfk,\nu} & 0
\end{pmatrix}, \qquad V^R_{\bfk,\nu} = \frac{1}{V_0^{-1}(\bfk) + \Pi^{R}_{\bfk,\nu}}.
\end{align}
In this formula, $V_0(\bfk) = {2 \pi e^2}/{|\bfk|}$ is the Coulomb interaction, $\Pi^{R}_{\bfk,\nu} = s \nu_0 \frac{D \bfk^2}{D \bfk^2-i \nu}$ is the retarded polarization operator, and $s=2$ counts the number of spin projections. With this definition, the advanced and the Keldysh components of the interaction are given by the formulas $V^A_{\bfk,\nu}=V^R_{\bfk,-\nu}$ and $V^K_{\bfk,\nu}= B_{\nu} (V^R_{\bfk,\nu} - V^A_{\bfk,\nu})$, respectively. Here, $B_\nu = \coth ({\nu}/{2 T})$ is the bosonic distribution function. It is worth noting that the polarization operator is not modified by the (weak) particle-hole asymmetry.

\subsection{Noninteracting case}

In order to set the stage for the calculation of interaction corrections to the dynamical part of the heat density-density correlation function $\chi_{kn}^{\op{dyn}}$ we briefly discuss the non-interacting case \cite{Fabrizio91,Schwiete21}.

\begin{figure}[h]
\includegraphics[width=4.0cm]{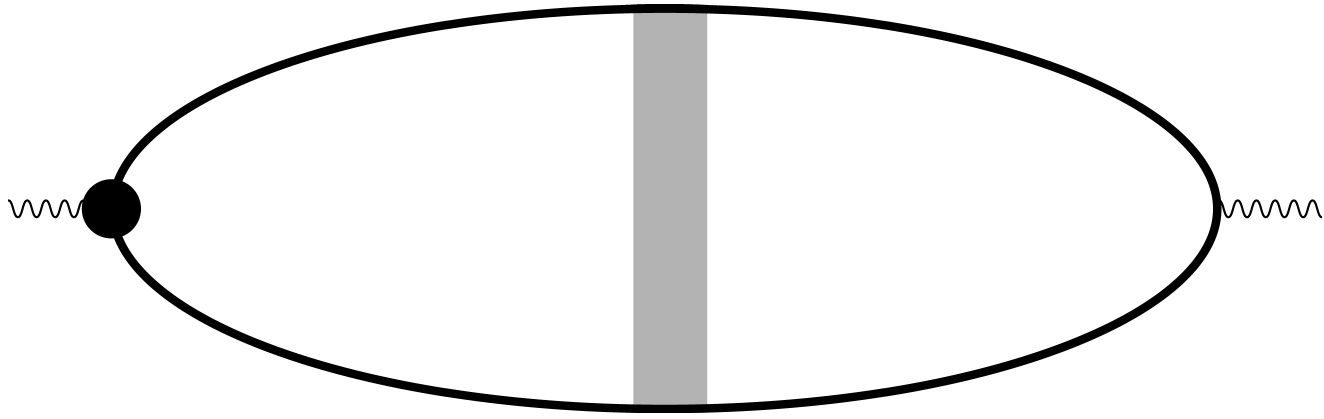}
\caption{The leading contribution to the dynamical part of the heat density-density correlation function in the noninteracting limit, $\chi_{kn,0}^{dyn}(\bfq,\omega)$. The shaded region is the diagrammatic representation of the diffuson $\mathcal{D}_{\bfq,\omega}^\eps$.}
\label{fig:chi0}
\end{figure}
The only term relevant for the non-interacting case originates from Eq.~\eqref{eq:chidyneps} (see Fig.~\ref{fig:chi0}) 
\begin{align}
\chi_{kn,0}^{\op{dyn}}(x_1,x_2) &= -\frac{i (\pi \nu_0)^2s}{4} \int_{\eps_i} e^{-i t_1(\eps_1-\eps_2)+i t_2(\eps_3-\eps_4)}\bar{\eps}_{12} \no\\
& \times \langle\tr[\gamma_2 \underline{\sigma_3 \hat{P}_{\eps_1 \eps_2}(\bfr_1)}] \tr[\gamma_1 \underline{\sigma_3 \hat{P}_{\eps_3 \eps_4}(\bfr_2)}]\rangle_0.
\end{align}
Using the contraction rules one finds
\begin{align}
\chi_{kn,0}^{\op{dyn}}(\bfq,\omg) &= - i \pi \nu_0 s \int_{\eps} \eps \Delta_{\eps,\omg} \mD_{\bfq,\omg}^{\eps}.
\end{align}
In the zero temperature limit $T\rightarrow 0$, the window function $\Delta_{\eps,\omg} = F_{\eps+\omg/2} - F_{\eps-\omg/2}$ with $F_\eps=\tanh\eps/2T$ restricts the frequency integral in $\eps$ to the interval $(-\omg/2,+\omg/2)$. This range broadens at finite temperatures. We recall that the label $\eps$ of the diffusion mode $\mD_{\bfq,\omg}^{\eps}$ signifies the frequency dependence due to particle-hole asymmetry. Without particle-hole asymmetry, $\chi_{kn}^{\op{dyn}}(\bfq,\omg)$ vanishes. After the expansion of $\mD_{\bfq,\omg}^{\eps}$ in $\eps$, one obtains
\begin{align}
\chi_{kn,0}^{\op{dyn}}(\bfq,\omg) &= c_0 T i \omg D_\eps' \bfq^2 \mD_{\bfq,\omg}^2.
\end{align}
Through a comparison with Eq.~\eqref{eq:chikndiff} with $\chi^{st}_{kn,0} =0$, \cite{Fabrizio91,Schwiete21} we identify $L=T c_0 D_\eps'$, which coincides with the result expected from Drude-Boltzmann transport theory.

\subsection{Interaction corrections}
\label{sec:intcorr}

This section is concerned with the calculation of interaction corrections to the dynamical part of the heat density-density correlation function originating from the combined effect of the long-ranged Coulomb interactions and disorder. The correlation function $\chi_{kn}$ vanishes in the absence of particle-hole asymmetry \cite{Schwiete21}. We can therefore proceed via a first order expansion of \eqref{eq:chidyneps} and \eqref{eq:chidynV} in the parameter $D'_\eps$. After inserting the $\zeta$-parameterization of Eq.~\eqref{parametrizationQ} for the matrix field $\hat{Q}$, Gaussian averages can be performed with the help of the contraction rule \eqref{eq:contraction2}. In order to capture the leading contribution in the dimensionless resistance $\rho$, it is sufficient to retain only contributions which require a one loop integration over momenta carried by the diffusion modes. 

Since the two frequency indices of the matrix $\hat{P}$ are inherited from electronic Green's functions, it is straightforward to construct representations of contributions to $\chi_{kn}^{\op{dyn}}$ in a conventional diagrammatic language \cite{abrikosov1975}. It is worth mentioning that numerical coefficients for individual diagrams may in general differ from those that would be obtained from a conventional Green'
s function approach to the calculation. The numerical coefficients do coincide, however, for the square root-even parameterization. As mentioned before, the $\zeta$ parameterization coincides with the square-root-even parameterization up to order $\hat{P}^4$ for the special case $\zeta=-1$. 

\begin{figure}[tb]
\includegraphics[width=8cm]{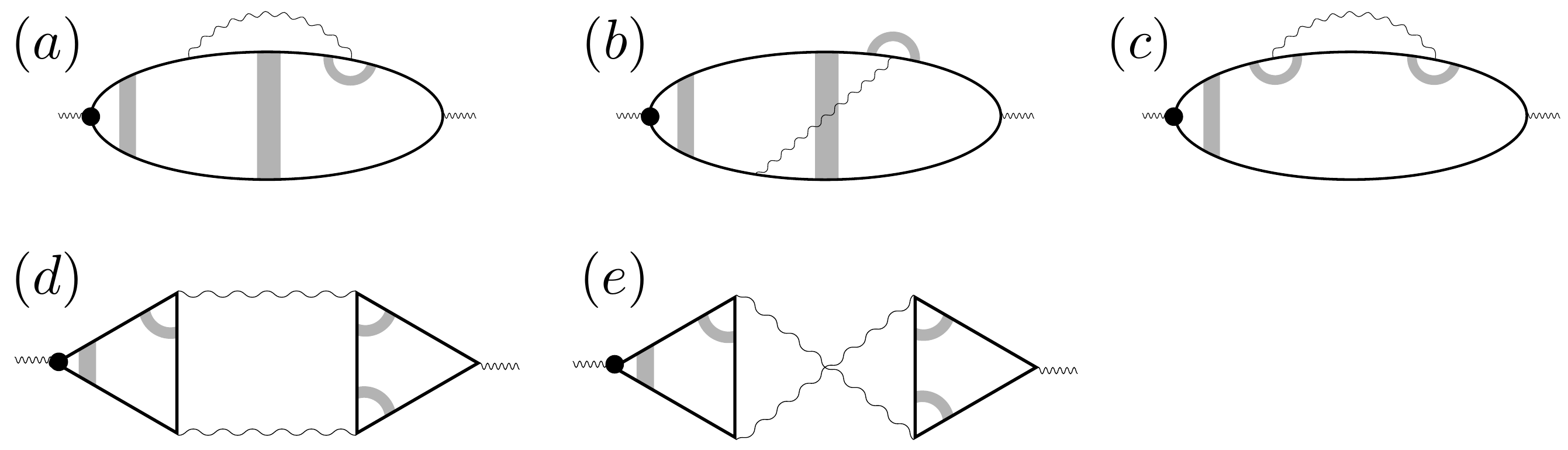}
\caption{Interaction corrections to the density vertex.}
\label{fig:diagrams_charge}
\end{figure}

\begin{figure}[tbh]
\includegraphics[width=8cm]{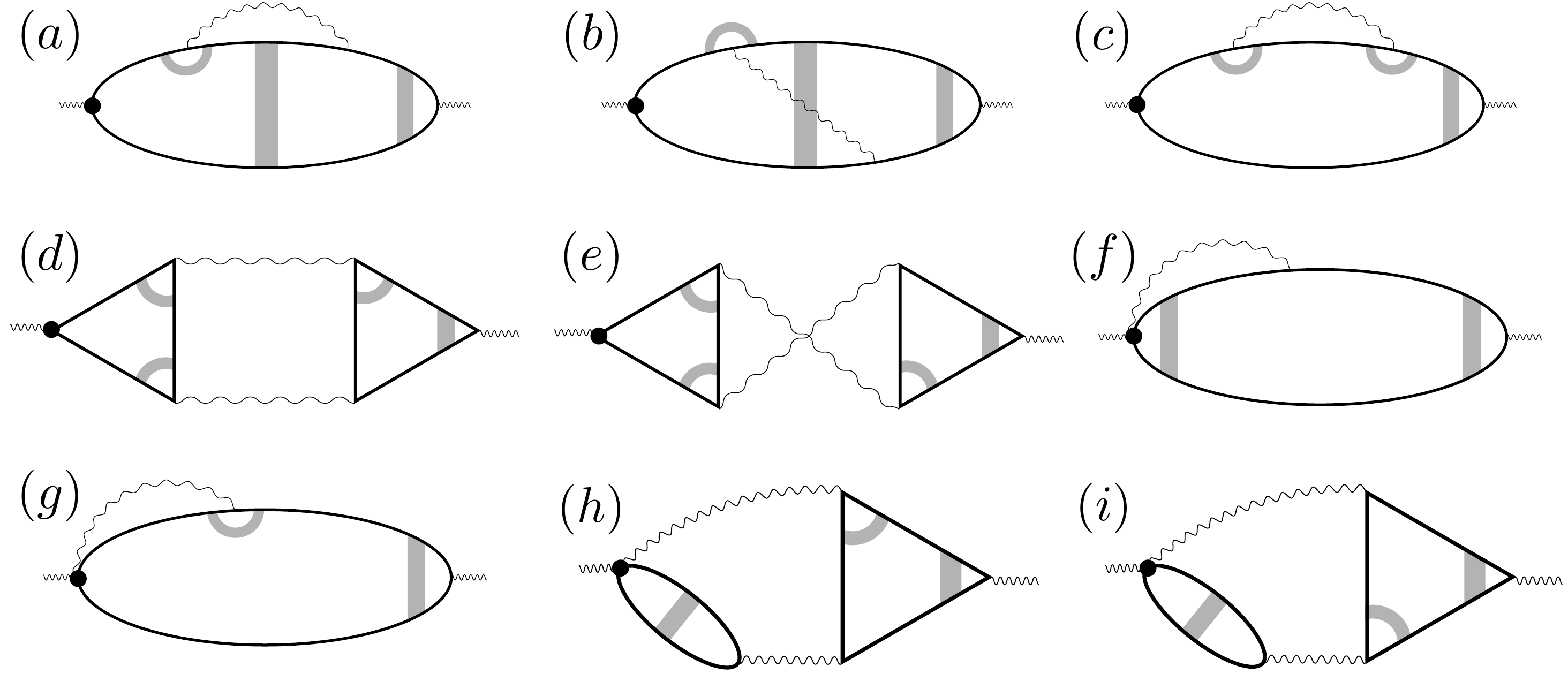}
\caption{Interaction corrections to the heat density vertex.}
\label{fig:diagrams_heat}
\end{figure}

Diagrams representing corrections to the diffuson for the heat density-density correlation function have already been displayed in Fig.~1 of paper I. To complete the calculation of $\chi^{dyn}_{kn}$, corrections to the density vertex and the heat density vertex also need to be accounted for. Diagrammatic representations for the charge vertex corrections and heat vertex corrections are displayed in Fig.~\ref{fig:diagrams_charge} and Fig.~\ref{fig:diagrams_heat}, respectively. In the NL$\sigma$M action, the gravitational potentials enter in combination with the frequency operator $\hat{\eps}$ and with the interaction term. As discussed in Sec.~\ref{sec:basic}, the heat density is therefore represented by two types of vertices in Fig.~\ref{fig:diagrams_heat}, one accompanied by frequency, the other one carrying an interaction line. Please note that in Fig.~1 of paper I, in Fig.~\ref{fig:diagrams_charge} and in Fig.~\ref{fig:diagrams_heat}, we omitted additional partner diagrams that can be obtained by a simple symmetrization of those already displayed. These partner diagrams are of course included into the calculation.

The parameter $D'_\eps$ can be extracted from different elements of each diagram, specifically from vertices, generalized Hikami boxes, and diffusons. One of the main advantages of working with the generalized NL$\sigma$M with particle-hole asymmetry compared to a conventional Green's function approach is that the dependence on $D'_\eps$ is explicit in the action, and the expansion in $D'_{\eps}$ is therefore straightforward. The diagrams in Fig.~1 of paper I, Fig.~\ref{fig:diagrams_charge} and Fig.~\ref{fig:diagrams_heat} are familiar from the calculation of interaction corrections to the density-density and heat density-heat density correlation functions, compare Ref.~\onlinecite{Schwiete16a}. As a result of the importance of particle-hole asymmetry for the calculation of $\chi_{kn}$, however, the frequency-dependence of the relevant contributions from each diagram is generally different from the above mentioned correlation functions.

Different types of logarithmic integrals are involved in the calculation of the interaction corrections to the heat density-density correlation function. These integrals can be classified according to the loop frequency ($\nu$) and loop momentum ($k$) regions that give the dominant contribution: 
\begin{enumerate}
\item For the first type of integrals, at least one of the two energies $|\nu|$ and $D\bfk^2$ lies in the RG energy interval $(T,1/\tau)$. Such integrals also appear for the density-density and heat density-heat density correlation function \cite{Finkelstein83,Castellani84,Castellani87,Schwiete16a}. Here, we label integrals of this type as $I_i$ and $I_i'$, see Appendix~\ref{sec:int}. 
\item The second type of integrals originates from frequencies constrained by the inequality $|\nu| < T$, while the momentum integration is determined by small momenta fulfilling the inequalities $|\nu|/(D \kappa_s) < k <\sqrt{|\nu|/D}$. In this case, the imaginary part of the dynamically screened Coulomb interaction is relevant and can be approximated as $\Im V^R_{\bfk,\nu}\approx -({1}/{2\nu_0})({\nu}/{D\bfk^2})$. The logarithm results from the $1/D\bfk^2$ singularity. Logarithmic integrals of this type will be denoted as $I_i^h$ (see Appendix~\ref{sec:int} for details). Such integrals are familiar from the calculation of the heat density-heat density correlation function \cite{Schwiete16a}. The momentum interval $|\nu|/(D\kappa_s) < k <\sqrt{|\nu|/D}$ is also responsible for the well-known double-logarithmic corrections to the tunneling density of states \cite{AltLee80} as well as other spurious corrections that appear in intermediate stages of the RG procedure for the disordered electron gas \cite{Finkelstein83,Castellani84}. We will also encounter such corrections in our calculation and the corresponding integral will be denoted as $I_1$. In contrast to the case of $I_1$, the integrals $I_i^h$ result in only a single logarithm. The key difference between the two cases is that for $I_i^h$ allowed frequencies $|\nu|$ are of the order of the temperature, while a double logarithm arises for $I_1$ because those frequencies can exceed the temperature.
\end{enumerate}

In addition to the corrections related to these two types of integrals, other corrections to the heat density-density correlation function arise during the calculation, which we will call $J$ terms. Unlike the corrections related to $I_i$, $I_i'$ and $I_i^h$, these additional corrections are not consistent with the general form of the correlation function discussed in Sec.~\ref{sec:chikngeneral}. These terms could, in principle, induce a gap in the diffuson, which would correspond to a violation of the conservation laws for charge and energy. However, the $J$ terms only arise as fragments of individual diagrams. As we will demonstrate in Appendix~\ref{sec:Jterms}, the $J$ terms cancel out when the contributions from all diagrams are taken into account.

This section is organized as follows. In Sec.~\ref{sec:charge_vertex}, we discuss the corrections to the charge vertex. We use this example to provide some details concerning the calculation with the generalized NL$\sigma$M with particle-hole asymmetry. Heat vertex corrections are the subject of Sec.~\ref{sec:heat_vertex}. Corrections to the diffuson are covered in Sec.~\ref{sec:diffuson}. Results for the interaction corrections to the thermopower are presented in Sec.~\ref{sec:thermo}. During our calculation, we encounter various logarithmic integrals, which are summarized and evaluated in Appendix~\ref{sec:int}. In Appendix~\ref{sec:Jterms}, we complete the calculation by addressing the cancellation of the $J$ terms. 

\subsubsection{Corrections to the charge vertex}
\label{sec:charge_vertex}

In this section, we discuss the calculation of the charge vertex corrections for the heat density-density correlation function in the presence of particle-hole asymmetry. We will label the vertex corrections according to the diagrams displayed in Fig.~\ref{fig:diagrams_charge}. For example, the charge ($\mathcal{C}$) vertex correction $\delta \chi_{kn,\mathcal{C}_a}$ corresponds to Fig.~\ref{fig:diagrams_charge}(a).

Charge vertex corrections are obtained by expanding the expression $\tr[\hat{\gamma_1} \ul{\delta \hat{X}}_{\eps_3 \eps_4}(\bfr_2)]$ in Eq.~\eqref{eq:chidyneps} beyond linear order in $\hat{P}$. Due to the relation $\delta \hat{X} = \delta \hat{Q} + (1/4 i) D_\eps' (\nabla \hat{Q})^2$, the corrections can be divided into two types. The first type involves the replacement $\delta \hat{X} \rightarrow \delta \hat{Q}$ in Eq.~\eqref{eq:chidyneps}. For these corrections, particle-hole asymmetry enters the calculation through the action $S_{\hat{\eta}=0,\hat{\phi}=0}$ in Eq.~\eqref{eq:Slin1}, i.e. through the frequency dependence of the diffuson in Eq.~\eqref{phDiffuson}, or through interaction vertices or Hikami boxes. The second type of corrections involves the replacement $\delta \hat{X} \rightarrow (1/4 i) D_\eps' (\nabla \hat{Q})^2$. In this case, particle-hole asymmetry enters through the explicit factor of $D_{\eps}'$. The averaging in Eq.~\eqref{eq:chidyneps} is then performed with respect to the conventional Finkel'stein action in the absence of particle-hole asymmetry. As a result, all diffusons become frequency-independent. We would like to stress that both types of corrections can contribute to the same diagram.

An evaluation of $\delta \chi_{kn,\mathcal{C}_a}$, $\delta \chi_{kn,\mathcal{C}_d}$, and $\delta \chi_{kn,\mathcal{C}_e}$ reveals that these corrections vanish. Specifically, $\delta \chi_{kn,\mathcal{C}_a}$ vanishes as it involves a frequency integration over a product of retarded functions. The corrections $\delta \chi_{kn,\mathcal{C}_d}$, and $\delta \chi_{kn,\mathcal{C}_e}$ can be seen to vanish as a result of the identity $V^{K}_\nu = B_\nu(V^R_\nu-V^A_\nu)$. The only remaining contributions read as
\begin{align}
&\delta \chi_{k n,\mathcal{C}_b}(\bfq,\omg) = 3 i \pi  \nu_0 s  D_{\eps}' \bfq^2\mD_{\bfq,\omg}^2  \int_{\eps} \eps^2 \Delta_{\eps,\omg} I_1(\eps)\no
\\
&\qquad - i \pi \nu_0 s  \mD_{\bfq,\omg} \int_{\eps} \eps^2 \Delta_{\eps,\omg} (3 I_1'(\eps) + I_3'(\eps)), \label{chargevertex}
\\
&\delta \chi_{k n, \mathcal{C}_c}(\bfq,\omg) =- \frac{3}{2}  i\pi \nu_0 s  (1+\zeta)D_{\eps}' \bfq^2\mD_{\bfq,\omg}^2  \int_{\eps} \eps^2 \Delta_{\eps,\omg} I_1(\eps) \no
\\
&\qquad + \frac{1 }{2} i  \pi \nu_0 s \mD_{\bfq,\omg} \int_{\eps} \eps^2 \Delta_{\eps,\omg} (3 (1+\zeta) I_1'(\eps) - 2 I_3'(\eps)). \no
\end{align}
These expressions contain certain logarithmic integrals $I_i(\eps)$ and $I'_i(\eps)$, which are listed in Appendix~\ref{sec:int}. The contribution from the integrals $I_1$ and $I_1'$ in $\delta \chi_{kn,\mathcal{C}_c}$ arises from the first type of correction introduced above Eq.~\eqref{chargevertex}, i.e. from the substitution $\delta \hat{X} \rightarrow \delta \hat{Q}$ in Eq.~\eqref{eq:chidyneps}. Interestingly, the charge vertex correction $\delta \chi_{kn,\mathcal{C}_c}$ does not appear in a conventional Green's function approach to the calculation \cite{Fabrizio91}. This can be understood as follows. The terms containing the integrals $I_1$ and $I_1'$ come with a factor of $(1+\zeta)$, which vanishes for the square-root even parameterization ($\zeta = -1$ in the one loop calculation). The remaining term in $\delta \chi_{kn,\mathcal{C}_c}$ is related to the integral $I_3'$. This term emerges due to the second type of substitution introduced above Eq.~\eqref{chargevertex}, $\delta \hat{X} \rightarrow (1/4 i) D_\eps' (\nabla \hat{Q})^2$, and does not vanish for $\zeta = -1$. As we will discuss at the end of the Sec.~\ref{sec:diffuson}, for the purpose of comparison with the conventional diagrammatic approach, this term should be grouped together with the Hikami box diagram that contribute to the correction to the diffuson. This is a recurrent theme that also applies to those contributions from $\delta \chi_{kn,\mathcal{H}_c}$, [Fig.~\ref{fig:diagrams_heat}(c)] and $\delta\chi_{kn,\mathcal{D}_b}$ [Fig.~1(b)], which arise from the substitution $\delta \hat{X} \rightarrow (1/4 i) D_\eps' (\nabla \hat{Q})^2$. 

We will now highlight typical manipulations involved in the evaluation of the interaction corrections to the correlation function using the example of $\delta \chi_{kn,\mathcal{C}_b}$. We start with the following expression, which is obtained after using the identity $B_{\eps-\eps'} (F_{\eps} - F_{\eps'}) = 1 - F_{\eps} F_{\eps'}$ and after discarding certain terms that vanish due to a frequency integration over retarded functions
\begin{widetext}
\begin{align}
\delta \chi_{k n,\mathcal{C}_b}(\bfq,\omg) &= \frac{\pi \nu_0 s}{2} \int_{\eps,\bfk,\nu} \eps \Delta_{\eps,\omega}\mD^{\eps}_{\bfq,\omega} \Big\{ F_{\eps_1-\nu}\mD^{\eps-\frac{\nu}{2}}_{\bfq+\bfk,\omega+\nu} \mD^{\eps+\frac{\omg-\nu}{2}}_{\bfk,\nu} -F_{\eps_2+\nu}\mD^{\eps+\frac{\nu}{2}}_{\bfq+\bfk,\omega+\nu} \mD^{\eps-\frac{\omg-\nu}{2}}_{\bfk,\nu}\bigg\} U^R_{\bfk, \nu} \no
\\
&- \frac{i \pi \nu_0 s D_\eps'}{4} \mD_{\bfq,\omg} \int_{\eps,\bfk,\nu} \eps \Delta_{\eps,\omg} (F_{\eps_2+\nu} + F_{\eps_1-\nu}) \mD_{\bfk,\nu} \mD_{\bfk+\bfq,\nu+\omg} (\bfk^2+\bfk \bfq) U^{R}_{\bfk,\nu}. \label{rawCharge}
\end{align}
Here, we introduced the notation: $\eps_1=\eps+\omg/2$, $\eps_2=\eps-\omg/2$. The first term originates from the substitution $\delta \hat{X} \rightarrow \delta \hat{Q}$ in Eq.~\eqref{eq:chidyneps}, while the second term comes from the substitution $\delta \hat{X} \rightarrow (1/4 i) D_\eps' (\nabla \hat{Q})^2$, as discussed above. A useful observation concerning both terms is that the window function $\Delta_{\eps,\omg} \approx \omg F'_{\eps}$ is of order $\omg$, so that $\omega$ or $\bfq$ can be neglected compared to other frequencies and momenta in the remaining parts of this expression. For the first term, the frequency dependent diffusons can be expanded as $\mD^{\eps}_{\bfp,\nu} \approx \mD_{\bfp,\nu} - \eps D_\eps' \bfp^2 \mD^{2}_{\bfp,\nu}$. In this way, we arrive at the following expression
\begin{align}
\delta \chi_{k n,\mathcal{C}_b}(\bfq,\omg) &=  \frac{\pi \nu_0 s D_\eps'}{2} \mD^{2}_{\bfq,\omega} \bfq^2 \int_{\eps,\bfk,\nu} \eps^2 \Delta_{\eps,\omg} (F_{\eps + \nu} - F_{\eps - \nu}) \mD_{\bfk,\nu}^2 U^R_{\bfk, \nu}\no
\\
&+\pi \nu_0 s  \mD_{\bfq,\omega} \int \eps^2 \Delta_{\eps,\omg} (F_{\eps+\nu} - F_{\eps-\nu}) \mD_{\bfk,\nu}^3 D_\eps' \bfk^2 U^R_{\bfk, \nu} \no
\\
&+\frac{\pi \nu_0 s}{2}  \mD_{\bfq,\omega} \int \eps \Delta_{\eps,\omg} (F_{\eps+\nu} + F_{\eps-\nu}) \nu \mD_{\bfk,\nu}^3 D_\eps' \bfk^2 U^R_{\bfk, \nu} \no
\\
&- \frac{i \pi \nu_0 s}{4} \mD_{\bfq,\omg} \int_{\eps,\bfk,\nu} \eps \Delta_{\eps,\omg} (F_{\eps+\nu} + F_{\eps-\nu}) \mD_{\bfk,\nu}^2 D_\eps' \bfk^2 U^{R}_{\bfk,\nu}.
\end{align}
\end{widetext}
We note that the third term is not logarithmically divergent. The other terms can be written in the form displayed in Eq.~\eqref{chargevertex} after introducing the logarithmic integrals $I_1(\eps)$, $I_1'(\eps)$ and $I_3'(\eps)$ defined in Appendix~\ref{sec:int}.

The expressions in Eq.~\eqref{chargevertex} can be further simplified. The window function $\Delta_{\eps,\omg}$ restricts the magnitude of relevant frequencies $\eps$ under the integrals to be of the order of the temperature $T$ or smaller. Since the integrals $I_i(\eps)$ and $I'_i(\eps)$ are functions of $\log[\max(\eps,T)]$, we can therefore use the approximation $\pi \nu_0 s \int_{\eps} \eps^2 \Delta_{\eps,\omg} I_i(\eps) = c_0 T I_i(\eps \rightarrow 0)$. For the sake of notational simplicity, we will denote $I_i(\eps\rightarrow 0)=I_i$ and $I_i'(\eps\rightarrow 0)=I'_i$. These integrals are listed and evaluated in Appendix~\ref{sec:int}. The charge vertex corrections in Eq.~\eqref{chargevertex} can now be written in the compact form
\begin{align}
&\delta \chi_{kn,\mathcal{C}}(\bfq,\omg)=c_0 Ti\omega \mathcal{D}_{\bfq,\omega}\times\no\\
&\qquad \times\left(\frac{3}{2} D_\eps'\bfq^2\mathcal{D}_{\bfq,\omega} (1-\zeta)I_1-\frac{3}{2} (1-\zeta)I_1' - 2 I_3'\right).\label{eq:chargevertex_Full}
\end{align}
The integral $I_1$ has a doubly logarithmic divergence. As in the calculation of $\chi_{kk}$ and $\chi_{nn}$, these double logarithms eventually cancel from $\chi_{kn}$ once the vertex corrections and corrections to the diffuson are combined. The logarithmic integrals $I_1'$ and $I_3'$ originate from the RG interval. We see that the correction to the charge vertex does not include any corrections from sub-thermal energies.

\paragraph*{Relation to the single-particle density of states:} It is instructive to note that the interaction corrections to the tunneling density of states in the presence of particle-hole asymmetry is \cite{Fabrizio91} 
\begin{align}
\delta\nu(\eps)/\nu_0=-3(I_1(\eps)+\eps I_1'(\eps))+\eps I_3'(\eps),\label{eq:deltanu}
\end{align}  
where the first part is the double logarithm and the terms proportional to $\eps$ arise due to particle-hole asymmetry and contain the same integrals $I_1'$ and $I_3'$ that we encountered for the charge vertex correction. In the NL$\sigma$M approach, the single particle density of states can be calculated as 
\begin{align}
\nu(\eps)=\frac{1}{2\pi i}\int_{\eps'}\frac{\partial \mathcal{Z}[\varphi]}{\delta \varphi(\bfr,\eps',\eps)},
\end{align}
where the dependence on the source field $\varphi$ is introduced into the formalism by adding the source term $S_\varphi=\pi\nu_0 \int_{\bfr}\tr[\sigma_+\hat{\sigma}_3\varphi \underline{X}]$ to the sigma model action ($\sigma_+$ projects onto the spin-up sector). The correction to the density of states naturally falls into two parts, $\delta \nu(\eps)=\delta \nu_1(\eps)+\delta\nu_2(\eps)$, where $\delta\nu_1(\eps)/\nu_0=-3(I_1(\eps)+\eps I_1'(\eps))$ is obtained from the substitution $X\rightarrow Q$ in $S_\varphi$, while $\delta\nu_2(\eps)/\nu_0= \eps I_3'(\eps)$ arises from the substitution $X\rightarrow  (1/4i)D_\eps'(\nabla \hat{Q})^2$ in $S_\varphi$. Correspondingly, the connection between the charge vertex correction and the density of states is
\begin{align}
\delta \chi_{kn,\mathcal{C}}(\bfq,\omg)&= \frac{i (1-\zeta)}{2 \nu_0} \int_{\eps} \eps \Delta_{\eps,\omg} \delta\nu_1(\eps) \mD^{\eps}_{\bfq,\omg}\no \\
& - \frac{2 i}{\nu_0} \mathcal{D}_{\bfq,\omega} \int_{\eps} \eps \Delta_{\eps,\omg} \delta\nu_2(\eps).
\end{align}
The correction $\delta\nu_1(\eps)$, which incorporates $I_1'$ in addition to the doubly logarithmic integral $I_1$, will cancel from the overall calculation once the vertex corrections and the corrections to the diffuson are combined. By contrast, such a cancellation does not occur for $\delta\nu_2(\eps)$.

\subsubsection{Corrections to the heat vertex}
\label{sec:heat_vertex}

This section is devoted to the calculation of the interaction corrections to the heat vertex, $\delta \chi_{kn,\mathcal{H}}$. The relevant diagrams are displayed in Fig.~\ref{fig:diagrams_heat}. Compared to the corrections to the charge vertex discussed in Sec.~\ref{sec:charge_vertex}, the heat vertex corrections have a more complicated structure. In addition to logarithmic corrections from the RG interval, which we will denote as $\delta \chi ^{>}_{kn,\mathcal{H}_m}$ for diagram $(m)$ in Fig.~\ref{fig:diagrams_heat}, logarithmic corrections {$\delta \chi ^{<}_{kn,\mathcal{H}_m}$} from the sub-thermal interval $(D \kappa_s^2 / T , T)$ arise for the heat vertex. A further complication is related to the corrections to the static part of the correlation function $\delta \chi_{kn}^{st}$. To understand this point better, it is instructive to revisit the general structure of the corrections to $\chi_{kn}$ displayed in Eq.~\eqref{eq:chiAtoE}, where $\delta \chi_{kn}^{A}(\bfq,\omg)$ is given by $2 i \omg \mD_{\bfq,\omg} \delta \chi_{kn}^{st}$. We have already obtained one half of this contribution as $- 2 c_0 T i \omg \mD_{\bfq,\omg} I_3' = i \omg \mD_{\bfq,\omg} \delta \chi_{kn}^{st}$ from the charge vertex corrections written in Eq.~\eqref{eq:chargevertex_Full}. The other half of this contribution, $i \omg \mD_{\bfq,\omg} \delta \chi_{kn}^{st}$, is expected to originate from the heat vertex correction. Unlike for the charge vertex, however, multiple diagrams contribute in the case of the heat vertex corrections. Moreover, extracting the final result requires a careful manipulations of the individual contributions. This is why we group these terms into a separate category. We denote them as $\delta \chi^*_{kn,\mathcal{H}_m}$ for diagram $(m)$ in Fig.~\ref{fig:diagrams_heat}. Following this scheme, the corrections to the heat vertex will be separated into three parts,
\begin{align}
\delta \chi_{k n, \mathcal{H}_m}(\bfq,\omg) &=  \delta \chi^>_{kn,\mathcal{H}_m}(\bfq,\omg)+ \delta \chi^<_{kn,\mathcal{H}_m}(\bfq,\omg)\no \\
& + \delta \chi^*_{kn,\mathcal{H}_m}(\bfq,\omg).
\end{align}

Before turning to the discussion of the individual contributions, let us note that $\delta \chi_{kn,\mathcal{H}_a}$ vanishes identically. The corrections from the RG interval read as
\begin{align}
&\delta \chi^>_{kn,\mathcal{H}_b}(\bfq,\omg) = i \omg c_0 T \times\no \\
&\qquad \times \left[3 D_{\eps}' \bfq^2 \mD_{\bfq,\omg}^2 I_1 -  \mD_{\bfq,\omg} (3 I_1' + I_3')\right], \label{heatsummary1}
\\
&\delta \chi^>_{kn,\mathcal{H}_c}(\bfq,\omg) =- \frac{1}{2} i \omg c_0 T \times \no\\
&\qquad \times \left[3 D_{\eps}' \bfq^2 \mD_{\bfq,\omg}^2 (1+\zeta) I_1-  \mD_{\bfq,\omg} (3 (1+\zeta) I_1' - 2 I_3')\right], \no
\\
&\delta \chi^>_{kn,\mathcal{H}_{f-g}}(\bfq,\omg) =- i \omg c_0 T \left[\mD_{\bfq,\omg}^2 D_{\eps}' \bfq^2 I_5 -  \mD_{\bfq,\omg} I_5'\right]. \no
\end{align}

One may observe that the results for the heat vertex corrections $\delta \chi^>_{kn,\mathcal{H}_b}$ and $\delta \chi^>_{kn,\mathcal{H}_c}$ agree with those for the corresponding charge vertex corrections, $\delta \chi_{k n, \mathcal{C}_b}$ and $\delta \chi_{k n, \mathcal{C}_c}$, respectively. As in the case of $\delta \chi_{k n, \mathcal{C}_c}$, the diagram related to $\delta \chi_{kn,\mathcal{H}_c}$, Fig.~\ref{fig:diagrams_heat}(c), is absent in diagrammatic perturbation theory. Since $\delta \chi^>_{kn,\mathcal{H}_c}$ is the only contribution to $\delta \chi_{kn,\mathcal{H}_c}$, the discussion presented in the context of $\delta \chi_{k n, \mathcal{C}_c}$ below Eq.~\eqref{chargevertex} carries over to $\delta \chi_{kn,\mathcal{H}_c}$ as well.

Next, we list the contributions arising from the sub-thermal interval
\begin{align}
\delta \chi^<_{kn,\mathcal{H}_b}(\bfq,\omg) &= \frac{1}{2} i \omg c_0 T \mD_{\bfq,\omg}^2 D_{\eps}' \bfq^2 I_1^h, \label{heatsummary2}
\\
\delta \chi^<_{kn,\mathcal{H}_{h-i}}(\bfq,\omg) &=- i \omg c_0 T \mD_{\bfq,\omg}^2 D_{\eps}' \bfq^2 I_2^h. \no
\end{align}
These type of corrections exist only for the heat vertex, not for the charge vertex. Due to the relation $I_1^h = 2 I_2^h$, however, the heat vertex corrections from the sub-thermal interval add up to zero as well.

The diagrams relevant for the correction $\delta \chi^*_{kn,\mathcal{H}}$ are displayed in Fig.~\ref{fig:diagrams_heat}(b) and Fig.~\ref{fig:diagrams_heat}(d)-(i). The analytic expressions read as
\begin{widetext}
\begin{align}
\delta \chi^*_{kn,\mathcal{H}_b}(\bfq,\omg) &= \frac{\pi \nu_0 s}{2} \mD_{\bfq,\omega} \int_{\eps,\bfk,\nu} \Delta_{\eps,\omg} (F_{\eps+\nu} - F_{\eps-\nu}) \nu^2 \mD_{\bfk,\nu}^3 D_\eps' \bfk^2 U^R_{\bfk,\nu}
\\
&- \frac{i \pi \nu_0 s}{4} \mD_{\bfq,\omg} \int_{\eps,\bfk,\nu} \Delta_{\eps,\omg} (F_{\eps+\nu} - F_{\eps-\nu}) \nu \mD_{\bfk,\nu}^2 D_\eps' \bfk^2 U^R_{\bfk,\nu}, \no
\\
\delta \chi^*_{kn,\mathcal{H}_{d-e}}(\bfq,\omg) &=- \frac{i \nu_0^2 s^2}{2} \omg \mD_{\bfq,\omg} \int_{\bfk,\nu} B_{\nu} \nu^3 \mD^4_{\bfk,\nu} D_\eps' \bfk^2 U^{R}_{\bfk,\nu} U^R_{\bfk, \nu}, \no
\\
\delta \chi^*_{kn,\mathcal{H}_{f-g}}(\bfq,\omg) &=- \frac{i \pi \nu_0 s}{8} \mD_{\bfq,\omg} \int_{\eps,\bfk,\nu} \Delta_{\eps,\omg} (F_{\eps+\nu} - F_{\eps-\nu}) \nu \mD^2_{\bfk,\nu} D_\eps' \bfk^2 (U^R_{\bfk,\nu} + U^A_{\bfk,\nu}) \no
\\
&- \frac{i \nu_0 s}{4} \omg \mD_{\bfq,\omg} \int_{\bfk,\nu} B_{\nu} \nu \mD^2_{\bfk,\nu} D_\eps' \bfk^2 (U^R_{\bfk,\nu} - U^A_{\bfk,\nu}), \no
\\
\delta \chi^*_{kn,\mathcal{H}_{h-i}}(\bfq,\omg) &=- \frac{\nu_0^2 s^2}{2} \omg \mD_{\bfq,\omg} \int_{\bfk,\nu} B_{\nu} \nu^2 \mD^3_{\bfk,\nu} D_\eps' \bfk^2 U^{R}_{\bfk,\nu} U^{R}_{\bfk,\nu}. \no
\end{align}
\end{widetext}
The sum of these contributions can be written as $\delta \chi^*_{kn,\mathcal{H}}=i \omg \mD_{\bfq,\omg} \delta \chi_{kn}^{st}$, where $\delta\chi_{kn}^{\op{st}}= - c_0 T I_4'$, see Appendix~\ref{sec:int}.

Now we can write the full expression for the heat vertex corrections
\begin{align}
\delta \chi_{kn,\mathcal{H}} (\bfq,\omega)=& c_0 Ti\omega \mathcal{D}_{\bfq,\omega}\times\no \\
&\times\left[D_\eps'\bfq^2\mathcal{D}_{\bfq,\omega} \left(\frac{3}{2} (1-\zeta)I_1 - I_5\right)\right.\no\\
&\left.-\frac{3}{2} (1-\zeta)I_1' + I_5' - 2 I_3' - I_4'\right]. \label{eq:heatvertex_Full}
\end{align}
Numerically, we find $I_4' = I_5' = 2 I_3'$, see Appendix~\ref{sec:int}. Before proceeding, we would like to draw the reader's attention to certain relations that exist between the heat vertex corrections calculated here and the heat vertex corrections for the heat density-heat density correlation function. Details are presented in Appendix~\ref{sec:connHeat}.

For later reference, we add the charge and heat vertex corrections for the heat density-density correlation function in Eqs.~\eqref{eq:chargevertex_Full} and \eqref{eq:heatvertex_Full} to find 
\begin{align}
&\delta \chi_{kn,\mathcal{C}}(\bfq,\omega) + \delta \chi_{kn,\mathcal{H}}(\bfq,\omega) = c_0 Ti\omega \mathcal{D}_{\bfq,\omega}\times \no\\
&\qquad \times\left[D_\eps'\bfq^2\mathcal{D}_{\bfq,\omega} \left[3 (1-\zeta)I_1 - I_5\right]\right.\no\\
&\qquad \left.- 3 (1-\zeta)I_1' + I_5' - 4 I_3' - I_4'\right]. \label{eq:vertex_Full}
\end{align}

\subsubsection{Corrections to the diffuson}
\label{sec:diffuson}

In this section, we address the corrections to the diffuson. Diagrams are shown in  Fig.~1 of paper I. We will first list the contributions from each individual diagram, and then present the combined result. We use the notation that $\delta \chi^<_{k n, \mathcal{D}_b}(\bfq,\omg)$ denotes a correction from the sub-thermal interval corresponding to the diagram shown in Fig.~1(b), while $\delta \chi^>_{k n, \mathcal{D}_b}(\bfq,\omg)$ denotes a correction from the RG interval. We will further classify the corrections for each diagram according to the decomposition presented in Eq.~\eqref{eq:chidecomp}. For example, $\delta \chi^{A>}_{k n, \mathcal{D}_b}(\bfq,\omg)$ denotes the RG interval correction from Fig.~1(b) that contributes to $\delta \chi^A_{kn}(\bfq,\omega)$. 
 
We start our discussion with the corrections that contribute to $\delta \chi^B_{kn}(\bfq,\omega)$, as they are directly related to the correction to the thermopower. From the RG interval we obtain
\begin{align}
\delta \chi^{B>}_{k n, \mathcal{D}_a}(\bfq,\omega) &= c_0 T i \omg \bfq^2 \mD_{\bfq,\omg}^2 (3 \zeta D_\eps' I_1 + D I_3'), \label{diff(B)}
\\
\delta \chi^{B>}_{k n, \mathcal{D}_b}(\bfq,\omega) &=- 2 c_0 T i \omg D \bfq^2 \mD_{\bfq,\omg}^2 I_3', \no
\\
\delta \chi^{B>}_{k n, \mathcal{D}_c}(\bfq,\omega) &=- c_0 T i \omg \bfq^2 \mD_{\bfq,\omg}^2 (3 D_\eps' I_1 + D I_D').  \no
\end{align}
Note that the integral $I_1$ is doubly logarithmic and will cancel out once all terms are accounted for. The contributions from Fig.~1(a) and 1(b) in the first two lines of Eq.~\eqref{diff(B)} and the $I_1$ integral in Fig.~1(c) were derived previously \cite{Fabrizio91}. Additionally, we found the logarithmic correction $I_D'$ originating from $\delta \chi^>_{k n, \mathcal{D}_c}$ (Fig.~1(c)), which is also relevant for the correction to the thermopower. This contribution was apparently not included in Ref.~\onlinecite{Fabrizio91}. A more detailed discussion of the role of $I_D'$ for the calculation of the thermopower will be presented in Sec.~\ref{sec:thermo}. 

The corrections from the sub-thermal interval have not been studied before. Here, we find
\begin{align}
\delta \chi^{B<}_{k n, \mathcal{D}_a}(\bfq,\omega) &= \frac{1}{4} c_0 T i \omg D_\eps' \bfq^2 \mD_{\bfq,\omg}^2 I^h_{1}, \label{diff(B)st}\\
\delta \chi^{B<}_{k n, \mathcal{D}_b}(\bfq,\omega) &=- \frac{1}{2} c_0 T i \omg D_\eps' \bfq^2 \mD_{\bfq,\omg}^2 I^h_{1},  \no
\\
\delta \chi^{B<}_{k n, \mathcal{D}_c}(\bfq,\omega) &= \frac{1}{4} c_0 T i \omg D_\eps' \bfq^2 \mD_{\bfq,\omg}^2 (I_1^h - 2 \tilde{I}_1^h + 2 I^h_{4}),  \no
\\
\delta \chi^{B<}_{k n, \mathcal{D}_d}(\bfq,\omega) &= \frac{1}{4} c_0 T i \omg D_\eps' \bfq^2 \mD_{\bfq,\omg}^2 (I_5^h - 6 \tilde{I}_1^h ),  \no
\\
\delta \chi^{B<}_{k n, \mathcal{D}_{e-h}}(\bfq,\omega) &=- 2 c_0 T i \omg D_{\eps}' \bfq^2 \mD_{\bfq,\omg}^2 I_2^h. \no
\end{align}
 
Next, we list the contributions to $\delta \chi^A_{kn}(\bfq,\omega)$
\begin{align}
\delta \chi^{A>}_{k n, \mathcal{D}_a}(\bfq,\omega) &=- 3 \zeta c_0 T i \omg \mD_{\bfq,\omg} I_1',
\\
\delta \chi^{A>}_{k n, \mathcal{D}_b}(\bfq,\omega) &= 3 c_0 T i \omg \mD_{\bfq,\omg} I_1'. \no
\end{align}
Just like the doubly logarithmic $I_1$ integral, the $I_1'$ integral is also part of the cancellation between the contributions from the vertex corrections and the corrections to the diffuson. We emphasize that both the charge vertex corrections and heat vertex corrections contribute to the terms $\delta \chi^{A}_{kn}(\bfq,\omega)$ and $\delta \chi^B_{kn}(\bfq,\omega)$ as well.
 
For the contributions to $\delta \chi^C_{kn}(\bfq, \omega)$ and $\delta \chi^D_{kn}(\bfq, \omega)$, in order to avoid unnecessarily complicated expressions, we state here the results of a summation over multiple diagrams rather than corrections corresponding to individual diagrams. For $\delta \chi^C_{kn}(\bfq, \omega)$, we get
\begin{align}
\delta \chi^{C>}_{k n, \mathcal{D}_{a-c}}(\bfq,\omega) &= \frac{1}{2} c_0 T \omg^2 \mD_{\bfq,\omg}^2 (4 I_3' + I_2' - 2 I_z'),
\\
\delta \chi^{C>}_{k n, \mathcal{D}_d}(\bfq,\omega) &= \frac{1}{2} c_0 T \omg^2 \mD_{\bfq,\omg}^2 (2 I_z' + I_6' - 3 I_2'). \no
\end{align}
As only corrections to the diffuson (Fig.~1 of paper I) contribute to $\delta \chi^C_{kn}(\bfq, \omega)$, the above expressions provide a useful check of the calculation. Indeed, the resulting combination of integrals (integrals are listed and evaluated in Appendix~\ref{sec:int}) fulfills the equality  $(4 I_3' + I_2' - 2 I_z') + (2 I_z' + I_6' - 3 I_2') = - 2 \partial_\mu z$. In view of Eq.~\eqref{eq:chiAtoE}, this is precisely what is required to confirm the relation $\delta \chi^{st}_{kn} = - c_0 T \partial_\mu z$. 

For the contributions to $\delta \chi^D_{kn}(\bfq,\omega)$, we find
\begin{align}
\delta \chi^{D>}_{k n, \mathcal{D}_{a-c}}(\bfq,\omega) &= 2 c_0 T \omg^2 D_\eps' \bfq^2 \mD_{\bfq,\omg}^3 I_z, \label{eq:Dterms}
\\
\delta \chi^{D>}_{k n, \mathcal{D}_d}(\bfq,\omega) &= c_0 T \omg^2 D_\eps' \bfq^2 \mD_{\bfq,\omg}^3 (3 I_2 - 2 I_z - I_6), \no
\\
\delta \chi^{D<}_{k n, \mathcal{D}_{e-h}}(\bfq,\omega) &= \frac{1}{2} c_0 T \omg^2 D_\eps' \bfq^2 \mD_{\bfq,\omg}^3 I_3^h. \no
\end{align}
As in the case of $\delta \chi^C_{kn}(\bfq, \omega)$, vertex corrections do not contribute to $\delta \chi^D_{kn}(\bfq, \omega)$. We therefore need to check if the expressions in Eq.~\eqref{eq:Dterms} add up to give  $\delta \chi_{kn}^D=-\omega^2 \bfq^2 \mD_{\bfq,\omega}^3 \left(\delta z - \frac{1}{2} I^h\right) L$ with the known expressions for $\delta z$ [Eq.~\eqref{eq:Iz}], $I^h$ [Eq.~\eqref{eq:Ih}] and $L$ [Sec.~\ref{sec:struc}]. This relation can indeed be confirmed with the help of the list of integrals stated in Appendix~\ref{sec:int}. Similar to the heat vertex corrections, there is also relationships between the corrections to the diffuson detailed here and the corresponding corrections found in the calculation of the heat density-heat density correlation function. We refer the interested reader to Appendix~\ref{sec:connDiff} for further details.
 
As for the contribution to $\delta \chi^E_{kn}$, we find 
\begin{align}
\delta \chi^{E>}_{k n, \mathcal{D}_c}(\bfq,\omega) &= 2 c_0 T i \omg D_{\eps}' D \bfq^4 \mD_{\bfq,\omg}^3 I_D.
\end{align}

{The integral $I_D$ has been evaluated in Appendix~\ref{sec:int}, Eq.~\eqref{eq:ID}. By identifying $I_D =- \delta D/D$ and $L= T c_0 D_\eps'$ we immediately recover the expression for $\delta \chi^E_{kn}$ in Eq.~\eqref{eq:chiAtoE}.}

The overall result for the corrections to the diffuson from the RG interval is
\begin{align}
&\delta \chi_{k n, \mathcal{D}}^{\op{dyn}>}(\bfq,\omega) = c_0 T  i\omg \mD_{\bfq,\omega}\times \no \\
&\times \bigg[3 (1 - \zeta) I_1' - \bfq^2 \mD_{\bfq,\omg} \left(3 (1 - \zeta) D_{\eps}' I_1 + D I_3' + D I_D'\right)\no \\
&- i \omg \mD_{\bfq,\omg} \left(I_6' /2+  2 I_3' - I_2'\right) \no
\\
&+ 2 D_{\eps}' D \bfq^4 \mD_{\bfq,\omg}^2 I_D - i \omg D_{\eps}' \bfq^2 \mD_{\bfq,\omg}^2 \left(3 I_2 - I_6\right)\bigg],\label{eq:diffusionFull}
\end{align}
and the corresponding result from the sub-thermal energy interval reads as
\begin{align}
\delta \chi_{k n, \mathcal{D}}^{\op{dyn}<}(\bfq,\omega) &= c_0 T i \omg \mD_{\bfq,\omg}  \times\no\\
&\times \left[D_{\eps}' \bfq^2 \mD_{\bfq,\omg} \left(-2 \tilde{I}_1^{h} + I_5^h/4 + I_{4}^h/2 - 2 I_2^h\right)\right.\no\\
&\left. - i \omg D_{\eps}' \bfq^2 \mD_{\bfq,\omg}^2 I_3^h/2 \right]. \label{eq:diffusionFullHeatLog}
\end{align}
We would like to remind the reader that the $J$ terms have not been included into the discussion here. We will treat them separately in Appendix~\ref{sec:Jterms}.

\subsubsection{Final results for the correlation function $\chi_{kn}$ and thermoelectric transport coefficient} 
\label{sec:thermofinal}

In Secs.~\ref{sec:charge_vertex}, \ref{sec:heat_vertex} and \ref{sec:diffuson}, we have discussed interaction corrections to the dynamical part of the heat density-density correlation function. The sum of all vertex corrections is displayed in Eq.~\eqref{eq:vertex_Full} and the results for the corrections to the diffuson in Eq.~\eqref{eq:diffusionFull} and Eq.~\eqref{eq:diffusionFullHeatLog}. As mentioned before, corrections to the static part of the correlation function $\delta \chi^{st}_{kn} = - c_0 T \partial_\mu z$ are known from previous works \cite{Fabrizio91,Schwiete21}. With the decomposition introduced in Eq.~\eqref{eq:chiAtoE}, the corrections can be summarized as follows
\begin{align}
\delta \chi^A_{kn}(\bfq,\omega) &= c_0 T i \omg \mD_{\bfq,\omg} (I_5' - 4 I_3' - I_4'), \label{eq:dchi}
\\
\delta \chi^B_{kn}(\bfq,\omega) &=- c_0 T i \omg \bfq^2 \mD_{\bfq,\omg}^2 \left(D I_3' + D I_D' + D_\eps' I_5\right. \no\\
&\left.+ 2 D_{\eps}' \tilde{I}_1^{h} + 2 D_{\eps}' I_2^h - D_{\eps}' I_5^h/4 - D_{\eps}' I_{4}^h/2\right), \no
\\
\delta \chi^C_{kn}(\bfq,\omega) &= c_0 T \omg^2 \mD_{\bfq,\omg}^2 \left(I_6'/2 +  2 I_3' - I_2'\right), \no
\\
\delta \chi^D_{kn}(\bfq,\omega) &= c_0 T \omg^2 \bfq^2 \mD_{\bfq,\omg}^3 \left(3 I_2 +  I_3^h/2 - I_6\right), \no
\\
\quad \delta \chi^E_{kn}(\bfq,\omega) &= 2 c_0 T i \omg D_\eps' D \bfq^4 \mD_{\bfq,\omg}^3 I_D. \no
\end{align}
A detailed account of all relevant integrals can be found in Appendix~\ref{sec:int}. Using specifically the results displayed in Eqs.~\eqref{eq:ID}, \eqref{eq:Iz} and \eqref{eq:Ih}, we find the following relations 
\begin{subequations}
\begin{align}
&I_5' - 4 I_3' - I_4' =- 2 \partial_\mu z, \label{eq:keya}\\
&\big(D I_3' + D I_D' + D_\eps' I_5 + 2 D_{\eps}' \tilde{I}_1^{h} + 2 D_{\eps}' I_2^h - D_{\eps}' I_5^h/4\no\\
 & - D_{\eps}' I_{4}^h/2\big) = D I_3' + D I_D' - D_\eps' \delta z + ({3}/{2}) D_\eps' I^h\label{eq:keyb}
\\
&I_6'/2 +  2 I_3' - I_2' = \partial_\mu z\\
&3 I_2 + I_3^h/2 - I_6 = - \left(\delta z -  I^h/2\right), \\
&D I_D = -\delta D, \label{eq:keye}
\end{align}
\end{subequations}
where $I^h=\rho\log(D\kappa_s^2/T)$. Once we identify $\delta L = - c_0 T (D I_3'+ D I_D' + D_\eps' I^h)$ in Eq.~\eqref{eq:keyb}, the relations \eqref{eq:dchi} and \eqref{eq:keya}-\eqref{eq:keye} are fully consistent with the perturbative expansion based on the phenomemological form of the correlation function, Eq.~\eqref{eq:chiAtoE}. We emphasize that we have identified the correction to $\delta z$ with the integral $I_5$ (and not with the integrals $I_D'$ and $I_3'$). In this way, all $I_i'$ integrals are associated with quantities that vanish in the absence of particle-hole asymmetry, such as $\delta L$ and $\delta \chi^{st}_{kn}=-c_0T\partial_\mu z$. A detailed discussion of the sub-thermal corrections contributing to Eq.~\eqref{eq:keyb} will be presented at the end of this section.

The term $\delta \chi_{kn,\mathcal{D}_a}$ plays a crucial role for the calculation of the thermoelectric transport coefficient. The associated diagram, Fig.~1(a) of paper I,  contains a diagrammatic block known as the Hikami box. In order to facilitate a comparison with the results of the diagrammatic analysis of Ref.~\onlinecite{Fabrizio91}, we contrast the Hikami box diagrams in the NL$\sigma$M formalism and in conventional diagrammatic many-body perturbation theory in Appendix~\ref{sec:hikami}.

We are now in a position to discuss the interaction corrections to $L$. To this end, it is convenient to write the result for $\delta L$ obtained from $\delta \chi^B_{kn}(\bfq,\omega)$ in Eq.~\eqref{eq:dchi} in combination with Eq.~\eqref{eq:keyb} in the form 
\begin{align}
\frac{\delta L}{L} &= \frac{\delta \alpha}{\alpha} =- \frac{D}{D_\eps'} (I_3'+I_D') - I^h. \label{connPaper}
\end{align}
Tracing back the origin of these corrections, one can see that the $I_3'$ term arises due to contributions from the generalized Hikami-box diagrams in Figs.~1(a) and 1(b), while $I_D'$ is related to the diagram in Fig.~1(c). While the correction related to $I_3'$ was previously obtained in Ref.~\onlinecite{Fabrizio91}, the RG-type contribution from $I_D'$ and the sub-thermal contribution from $I^h$ were absent in this work. 

In paper I, we employed a somewhat different notation for the individual terms in Eq.~\eqref{connPaper} in order to streamline the presentation. From the definition of $I_3'$ in Eq.~\eqref{eq:I3_def}, we see that $I_3' = - \delta D_\eps^{(2)}/(\eps D)$, where $\delta D_\eps^{(2)}$ was introduced in paper I. Similarly, we can identify $I_D' = - (\delta D_\eps^{(1)}-\delta D)/(\eps D)$. Now we can use the relations $\delta D_\eps^{(1)} - \delta D= - \frac{1}{2} \eps D_\eps' I$ and $\delta D_\eps^{(2)}= - \frac{1}{4} \eps D_\eps' I$ derived in the paper I, and the well-known correction to the conductivity \cite{AltLee80} $\delta \sigma/\sigma=-I$ with $I=\rho\log1/T\tau$, to arrive at the main results of our study
\begin{align}
\frac{\delta \alpha}{\alpha} &=\frac{3}{4}\frac{\delta \sigma}{\sigma}-I^h. \label{eq:deltaAlpha}
\end{align}
We refer the reader to Sec.~\ref{sec:thermo} for a further discussion of this result.

To conclude this section, we return to the discussion of the sub-thermal corrections to $\delta \chi^B_{kn}(\bfq,\omega)$ in connection with Eqs.~\eqref{eq:dchi} and \eqref{eq:keyb}. The full list of contributions reads as
\begin{align}
\delta \chi^{B<}_{k n, \mathcal{D}_a}(\bfq,\omega) &= \frac{1}{4} c_0 T i \omg D_\eps' \bfq^2 \mD_{\bfq,\omg}^2 I^h_{1}, \\
\delta \chi^{B<}_{k n, \mathcal{D}_b}(\bfq,\omega) &=- \frac{1}{2} c_0 T i \omg D_\eps' \bfq^2 \mD_{\bfq,\omg}^2 I^h_{1}, \no
\\
\delta \chi^{B<}_{k n, \mathcal{D}_c}(\bfq,\omega) &= \frac{1}{4} c_0 T i \omg D_\eps' \bfq^2 \mD_{\bfq,\omg}^2 (I_1^h - 2 \tilde{I}_1^h + 2 I^h_{4}),\no\\
\delta \chi^{B<}_{k n, \mathcal{D}_d}(\bfq,\omega) &= \frac{1}{4} c_0 T i \omg D_\eps' \bfq^2 \mD_{\bfq,\omg}^2 (I_5^h - 6 \tilde{I}_1^h ),  \no
\\
\delta \chi^{B<}_{k n, \mathcal{D}_{e-h}}(\bfq,\omega) &=- 2 c_0 T i \omg D_{\eps}' \bfq^2 \mD_{\bfq,\omg}^2 I_2^h. \no
\end{align}
We can group all the diagrams with a horizontal interaction line [Fig.~1(a), (b) and (c)] together and arrive at the expressions
\begin{align}
\delta \chi^{B<}_{k n, \mathcal{D}_{a-c}}(\bfq,\omega) &= \frac{1}{4} c_0 T i \omg D_\eps' \bfq^2 \mD_{\bfq,\omg}^2 (2 I^h_{4} - 2 \tilde{I}_1^h), \\
 \delta \chi^{B<}_{k n, \mathcal{D}_d}(\bfq,\omega) &= \frac{1}{4} c_0 T i \omg D_\eps' \bfq^2 \mD_{\bfq,\omg}^2 (I_5^h - 6 \tilde{I}_1^h ),  \no
\\
\delta \chi^{B<}_{k n, \mathcal{D}_{e-h}}(\bfq,\omega) &=- 2 c_0 T i \omg D_{\eps}' \bfq^2 \mD_{\bfq,\omg}^2 I_2^h. \no
\end{align}
The contribution from $\delta \chi^{B<}_{k n, \mathcal{D}_{a-c}}(\bfq,\omega)$ exactly vanishes due to the relation $I^h_{4} = \tilde{I}_1^h$, so the diagrams in Fig.~1(a), (b) and (c) do not give a net contribution to the sub-thermal correction to the thermopower. We can read off the results for the remaining integrals from Eq.~\eqref{eq:Ih}, $(I_5^h - 6 \tilde{I}_1^h )/4 - 2 I_2^h = - (3/2) I^h$. This identity allows us to conclude that the sub-thermal correction to thermoelectric transport coefficient is $\delta L_< =-I^h$. This completes the derivation of the interaction corrections to $L$ and $\alpha$.

\section{Thermopower}
\label{sec:thermo}

We can use the result for the thermoelectric transport coefficient in Eq.~\eqref{eq:deltaAlpha} and the relation $S = \alpha/\sigma = e L/(\sigma T)$, to obtain the following expression for the interaction correction to the thermopower
\begin{align}
\frac{\delta S}{S}=-\frac{1}{4}\frac{\delta \sigma}{\sigma}-I^h. \label{eq:deltaS}
\end{align}
where we repeat $I^h = \rho \log(D \kappa_s^2/T)$ and $\delta\sigma/\sigma=-\rho \log 1/T\tau$.
A few comments are in order here.

1.~As discussed in the previous sections, both real and virtual processes contribute to the corrections to the thermoelectric transport coefficient $\alpha$ and the thermopower $S$. Specifically, the first term on the right hand side of Eq.~\eqref{eq:deltaS} originates from virtual processes, and the second term is the result of real processes.

\begin{figure}[b]
\includegraphics[width=8cm]{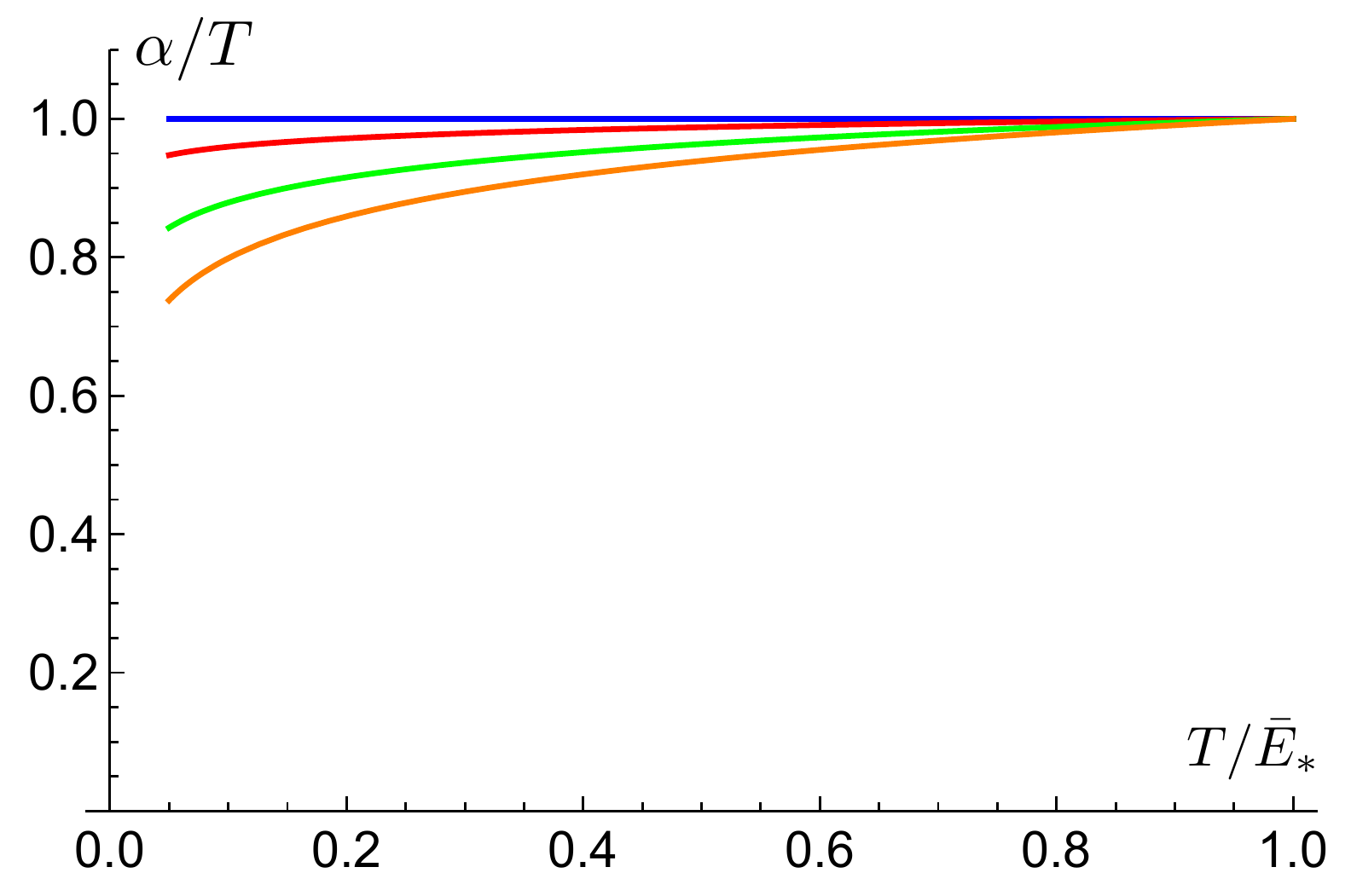}
\caption{The ratio $\alpha/T$ in units of $\alpha(\bar{E}_*)/\bar{E}_*$ according to Eq.~\eqref{eq:da_a}. From top to bottom the parameters are $\rho=0,0.01,0.03,0.05$. The case $\rho=0$ corresponds to the classical case, for which interaction corrections are absent. For the sake of simplicity, we set $\bar{E}_*=D\kappa_s^2=1/\tau$.}
\label{fig:diagrams_alpha_T}
\end{figure}

2.~The logarithmic corrections to $S$ from the RG interval $(\delta \sigma/\sigma<0)$ and sub-thermal energy interval $(I^h>0)$ are opposite in sign, and as such compete with each other. Overall, the latter contribution dominates and the sign of the correction to the thermopower is negative
\begin{align}
\frac{\delta S}{S}=-\frac{3}{4}\rho\log\frac{\bar{E}}{T}<0, \quad \bar{E}=(D\kappa_s^2)^{4/3}\tau^{1/3}.\label{eq:dSoverS}
\end{align}
By contrast, the corrections to $\alpha$ from the RG and sub-thermal energy intervals are both negative, see Eq.~\eqref{eq:deltaAlpha}, so that 
\begin{align}
\frac{\delta \alpha}{\alpha}=-\frac{7}{4}\rho\log\frac{\bar{E}_*}{T}<0,\quad \bar{E}_*=(D\kappa_s^2)^{4/7}\tau^{-3/7},\label{eq:da_a}
\end{align}
and the overall effect of the interaction corrections is stronger for $\alpha$ than for $S$. For both Eq.~\eqref{eq:dSoverS} and $\eqref{eq:da_a}$ it is assumed that $T<\min(D\kappa_s^2,1/\tau)$. We illustrate the result for the interaction corrections to $\alpha$ in Fig.~\ref{fig:diagrams_alpha_T}.

3.~We can compare our result to the one derived in Ref.~\onlinecite{Fabrizio91}, namely $\delta S/S = - (3/4) \delta \sigma /\sigma$. Besides the absence of the correction from the sub-thermal energy interval, we notice that the coefficient for the corrections from the RG interval obtained in Ref.~\onlinecite{Fabrizio91} is also different from that in Eq.~\eqref{eq:deltaS}. We have traced the mismatch of the coefficient of the RG-interval corrections to a specific term (the term containing the $I_D'$ integral in Eq.~\eqref{diff(B)}), which appears to be missing in the previous study. 

4.~Application of the developed theory to Si-MOSFETs requires the inclusion of the valley degree of freedom $n_v=2$, see Ref.~\cite{Punnoose01}. The results of Drude-Boltzmann theory for the transport coefficients $\sigma$ and $\alpha$ are then proportional to $n_v$. The one-loop interaction corrections to these coefficients, $\delta \sigma$ and $\delta \alpha$, however, are independent of the valley degree of freedom. This is because for all relevant diagrams the number of closed loops corresponds to the number of screened interaction lines. While each loop is proportional to $n_v$, the screened Coulomb interaction line is inversely proportional to $n_v$, so that all factors of $n_v$ cancel out. This reasoning applies both for the corrections from real and from virtual processes. The relative deviations $\delta \sigma/\sigma$, $\delta \alpha/\alpha$, and $\delta S/S$ are therefore proportional to $n_v^{-1}$.

5.~The thermoelectric figure of merit $z$ is a measure for the efficiency with which a system is able to convert heat into electrical energy and vice versa. The figure of merit is defined by the relation
\begin{align}
zT = \sigma S^2 T / \kappa = S^2/\mathcal{L},
\end{align}
where $\mathcal{L}=\kappa/\sigma T$ is the Lorenz number. How do the interaction corrections affect the figure of merit for the disordered electron gas with Coulomb interactions? By combining the well-known correction to the Lorenz number $\delta \mathcal{L}/\mathcal{L}_0= I^h/2$, which was obtained in Refs.~\onlinecite{Raimondi04,Niven05,Catelani05,Michaeli09,Schwiete16a,Schwiete16b} and reflects the violation of the Wiedemann-Franz law, with the correction to the thermopower displayed in Eq.~\eqref{eq:deltaS}, we obtain
\begin{align}
\frac{\delta zT}{zT} &= -\frac{1}{2}\frac{\delta \sigma}{\sigma}- \frac{5}{2} I^h. \label{eq:deltazT}
\end{align}
Similar to Eqs.~\eqref{eq:dSoverS} and \eqref{eq:da_a}, we can cast the result in a compact form
\begin{align}
\frac{\delta zT}{zT} &= -2 \rho \log\frac{\bar{E}_{**}}{T},
\end{align}
where $\bar{E}_{**}= (D \kappa_s^2)^{5/4} \tau^{1/4}$.
We see that the contribution from real processes (processes originating from sub-thermal energy interval) dominates compared to virtual processes. Therefore, the overall effect of the (Coulomb) interaction corrections on the figure of merit is a logarithmic suppression.

\section{Conclusion}
\label{sec:conclusion}

This manuscript contains a detailed account of the one-loop interaction corrections to the thermoelectric transport coefficient $\alpha$ and thermopower $S$ in the two-dimensional disordered electron gas. The main results of our study are displayed in Eqs.~\eqref{eq:deltaAlpha} and \eqref{eq:deltaS}, illustrated in Fig.~\ref{fig:diagrams_alpha_T} and discussed in Sec.~\ref{sec:thermo}. Our results for $\alpha$ and $S$ differ in magnitude and in sign from the previous theoretical work of Ref.~\onlinecite{Fabrizio91}. There are two reasons for this discrepancy. First, we demonstrated that real processes, which were not considered in Ref.~\onlinecite{Fabrizio91}, contribute to the logarithmic corrections to $\alpha$ and $S$. Second, we obtain a different coefficient for the corrections from virtual processes compared to Ref.~\onlinecite{Fabrizio91}. The origin of this mismatch is discussed in Secs.~\ref{sec:diffuson} and \ref{sec:thermofinal}. 

In principle, one may expect weak localization effects to give rise to quantum corrections of similar magnitude and temperature dependence as the interaction corrections. It is predicted theoretically, however, that weak localization does not affect the diffusion contribution to the thermoelectric transport coefficient $\alpha$ at first order in $\rho$ \cite{Castellani88a}. The temperature-dependence of $\alpha/T$ is therefore expected to be dominated by the interaction corrections calculated in this manuscript. By contrast, weak localization can still contribute to the thermopower $S=\alpha/\sigma$ through the correction to $\sigma$. This observation has important consequences for the magnetic field-dependence of the transport coefficients $\alpha$ and $S$. Interaction corrections caused by the long-range part of the Coulomb interaction are only weakly magnetic field-dependent \cite{Altshuler85}. This statement also applies to the interaction corrections to $\alpha$ calculated in this manuscript. By contrast, weak localization corrections are easily destroyed by the orbital contribution of the magnetic field. This strong magnetic field-dependence carries over to the thermopower $S$ through its dependence on $\sigma$.

Measurements of the thermopower of disordered electron systems are complicated by the presence of two competing effects: phonon drag and diffusion. Our work is concerned with the calculation of the diffusion component of the thermopower only, which is expected to be the dominant effect at low temperatures. Please note that our work addresses a different transport regime compared to Refs.~\onlinecite{Dolgopolov11,Gold11}. The focus of these references is the diffusion thermopower in the ballistic regime $T\gg 1/\tau$, a regime where the interaction corrections to the electric conductivity have a characteristic linear temperature dependence \cite{Gold86,DasSarma99,Zala01}.

The diffusion thermopower has been the subject of several experimental studies \cite{Fletcher01,Rafael04,Mokashi12}. In Ref.~\onlinecite{Rafael04}, weak localization effects have been investigated. In contrast to interaction corrections, weak localization corrections display a pronounced magnetic field dependence, which in principle allows to distinguish these two effects experimentally. While the leading diffusion contribution to the thermopower ($S \propto T$) could be identified in this experiment, the accuracy was unfortunately not sufficient to distinguish between a $T$ and a $T \log T$ dependence. Such a distinction would be necessary for a measurement of weak localization as well as of interaction corrections. In Ref.~\onlinecite{Fletcher01} and Ref.~\onlinecite{Mokashi12}, the diffusion thermopower was measured close to the metal-insulator transition in Si-MOSFET devices and observed to display a critical density-dependence near the transition. However, the temperature-dependence of quantum corrections was unfortunately not analyzed in detail.

This work could be generalized in several directions. We performed all calculations using a quadratic dispersion in two spatial dimensions. In this case, the origin of the particle-hole asymmetry is the non-constancy of the electron velocity, while the density of states and scattering time can be taken as frequency-independent. To make the technique applicable to a wider range of systems, a generalization to other electronic dispersions and different dimensionalities would be necessary. Technically, such problems could be addressed either by a further generalization of the NL$\sigma$M to include the effects of a non-constant density of states and scattering time, or by using conventional diagrammatic techniques. For example, using a quadratic dispersion in three dimensions would allow us to model a thin disordered films, for which diffusion modes are effectively two-dimensional as a result of their quantization in the transverse direction. An enhancement of the thermopower is known to occur near a Lifshitz transition \cite{Lifshitz60,Varlamov21}. The origin of this effect lies in the particle-hole asymmetry related to the strong frequency-dependence of the density of states and elastic scattering time as caused by the associated change of the spectrum near the transition. In the two-dimensional electron gas a Lifshitz transition occurs when the Fermi energy crosses the next size-quantization level \cite{Ablyazov91,Blanter92}. An application of the developed formalism to the calculation of quantum corrections to the thermopower near the Lifshitz transition would therefore also require a generalization to non-constant density of states and scattering time. A topic of great current interest is transport in quantum critical metals \cite{Brando16}, in which strong interactions lead to the inapplicability of the Fermi-liquid paradigm. A number of recent works have studied the influence of spatial randomness in the disorder potential \cite{Nosov20,Wu22,Guo22,Patel23} and in the Yukawa-coupling \cite{Aldape22,Patel23} on the transport properties of such non-Fermi liquids. In particular, Ref.~\cite{Patel23} pointed out the potential relevance of a combination of these two sources of randomness for the description of strange metal phases. Interaction corrections have also been the subject of investigation in this context \cite{Wu22}. So far, these studies mainly focused on electric transport; it would certainly be interesting to explore thermal and thermoelectric transport properties as well.

The present work can also be viewed as a first step towards developing a generalization of the two-parameter scaling theory of the disordered electron liquid to thermoelectric transport phenomena. The two-parameter scaling theory \onlinecite{Punnoose01} was developed for the description of the metallic side of the metal-insulator transition (for a recent review see \cite{Finkelstein23}). A generalization to thermoelectric transport would necessarily require an RG analysis of the NL$\sigma$M with particle-hole asymmetry. This analysis should be supplemented with the calculation of the interaction corrections from the sub-thermal energy interval with the help of the renormalized theory. Such a program has recently been implemented for thermal transport in Refs.~\onlinecite{Schwiete14b,Schwiete16a,Schwiete16b}.  

\section{Acknowledgments} The authors would like to thank A.~M.~Finkel'stein and T.~Micklitz for discussions, and K.~Michaeli for collaboration during the early stages of this work. This work was supported by the National Science Foundation (NSF) under Grant No. DMR-1742752.  

\appendix
\section{General form of the heat density-density correlation function}
\label{app:gen}
In this section, we present a derivation of Eq.~\ref{eq:chikndiff}, the heat density-density correlation function $\chi_{kn}$ in the diffusive limit. To find this correlation function, we consider long-wavelength fluctuations of the heat density $\delta k=\delta \eps-\mu \delta n=T\delta s$ in response to a smooth and slowly varying external potential $\varphi$. Here, $\delta \eps$, $\delta n$ and $\delta s$ denote fluctuations of the energy density, particle density, and entropy density, respectively \cite{Charge_Remark}. The correlation function $\chi_{kn}$ can then be found from the linear response relation $\delta k(\bfq,\omega)=\chi_{kn}(\bfq,\omega)\varphi(\bfq,\omega)$ for the Fourier components with momentum $\bfq$ and frequency $\omega$. It will be convenient to include the density-density correlation function $\chi_{nn}$ into the discussion as well, based on the relation $\delta n(\bfq,\omega)=\chi_{nn}(\bfq,\omega)\varphi(\bfq,\omega)$.

The local conservation laws for the particle number and energy result in the two continuity equations
\begin{align}
\left(\ba{cc} \partial_t \delta n\\ T\partial_t \delta s\ea\right)=-\nabla\cdot \left(\ba{cc}   {\bf j}\\ {\bf j}_k\ea\right),\label{eq:continuity}
\end{align}
where ${\bf j}$ is the particle current and ${\bf j}_k$ is the heat current. In the presence of an external potential $\varphi$, the currents are connected to gradients of the temperature and the chemical potential, and to the electric field ${\bf E}=-\nabla\varphi$ through the thermoelectric transport coefficients 
\begin{align}
\left(\begin{array}{cc} {\bf j}\\ {\bf j}_k\end{array}\right)=\left(\begin{array}{cc} \sigma&\alpha\\ \alpha T&\tilde{\kappa}\end{array}\right)\left(\ba{cc} {\bf E}-\nabla \mu\\ -\nabla T\ea\right)\label{eq:coefficients}.
\end{align}
In order to close the system of equations for the density fluctuations, we introduce a matrix of thermodynamic susceptibilities as
\begin{align}
\nabla \left(\ba{cc} \delta n\\ \delta s\ea\right)=\left(\ba{cc} \chi&\zeta\\\zeta&c_\mu/T\ea\right)\nabla \left(\ba{cc} \mu\\ T\ea\right).\label{eq:susceptibilities}
\end{align}
Here, the susceptibilities are obtained as second derivatives of the grand canonical potential density, $\omega=\eps-Ts-\mu n$, 
\begin{gather}
c_\mu\equiv-T\partial^2_T \omega=T\partial_T s,\quad  \chi\equiv-\partial^2_\mu \omega=\partial_\mu n,\\
 \zeta\equiv-\partial_\mu\partial_T\omega=\partial_T n=\partial_\mu s.  \nonumber
\end{gather}
By combining the relations in Eqs.~\eqref{eq:continuity}, \eqref{eq:coefficients} and \eqref{eq:susceptibilities}, we find a diffusion equation of the form
\begin{align}
&\partial_t\left(\ba{cc} \delta n\\ \delta s\ea\right)=\hat{D}\nabla^2\left(\ba{cc} \delta n+\varphi\chi\\ \delta s+\varphi \zeta\ea \right),\label{eq:diffequation}
\end{align}
where the diffusivity matrix is defined as 
\begin{align}
&\hat{D}=\left(\begin{array}{cc} \sigma&\alpha\\ \alpha &\tilde{\kappa}/T\end{array}\right)\left(\ba{cc} \chi&\zeta\\\zeta&c_\mu/T\ea\right)^{-1}.
\end{align}
In order to cast the diffusion equation in the form of Eq.~\eqref{eq:diffequation},
we used the relation $\hat{D}^{-1}(\sigma,\alpha)^t=(\chi,\zeta)^t$. A closely related discussion can be found in Refs.~\onlinecite{hartnoll2015,blake2017}. Written in Fourier components, the resulting expressions for $\chi_{nn}$ and $\chi_{kn}$ read as \begin{align}
\left(\ba{cc} \chi_{nn}(\bfq,\omega)\\ \frac{1}{T}\chi_{kn}(\bfq,\omega)\ea\right)=-\frac{\hat{D}\bfq^2}{\hat{D}\bfq^2-i\omega}\left(\ba{cc} \chi \\  \zeta \ea\right).\label{eq:standardform}
\end{align}
We can immediately read off the static parts of the correlation functions 
\begin{align}
\chi^{st}_{nn}&=\lim_{\bfq\rightarrow 0}\lim_{\omega\rightarrow 0}\chi_{nn}(\bfq,\omega)=-\chi,\\
\chi^{st}_{kn}&=\lim_{\bfq\rightarrow 0}\lim_{\omega\rightarrow 0}\chi_{kn}(\bfq,\omega)=-\zeta T,\nonumber
\end{align}
as well as the important constraints $\lim_{\omega\rightarrow 0}\lim_{\bfq\rightarrow 0}\chi_{nn}(\bfq,\omega)=\lim_{\omega\rightarrow 0}\lim_{\bfq\rightarrow 0}\chi_{kn}(\bfq,\omega)=0$, which reflect the conservation laws for energy and particle number.

For the subsequent discussion, it will be useful to rewrite Eq.~\eqref{eq:standardform} as 
\begin{align}
\left(\ba{cc} \chi_{nn}(\bfq,\omega)\\ \frac{1}{T}\chi_{kn}(\bfq,\omega)\ea\right)&=-\frac{\det[\hat{D}]\bfq^4-i\omega\bfq^2 \hat{D}}{\det[\hat{D}]\bfq^4-i\omega\tr[\hat{D}]\bfq^2-\omega^2}\left(\ba{cc} \chi\\\zeta\ea\right).\label{eq:factor}
\end{align}
An explicit calculation results in the following expressions for the determinant and trace of the diffusivity matrix
\begin{align}
\det[\hat{D}]=\frac{\sigma}{\chi}\frac{\kappa}{c_n},\quad \tr[\hat{D}]=\frac{\sigma}{\chi}+\frac{\kappa}{c_n}+\frac{T}{c_n\chi^2\sigma}(\zeta\sigma-\chi\alpha)^2,\label{eq:dettr}
\end{align}
where $c_n=c_\mu-{T\zeta^2}/{\chi}$ is the specific heat at constant density, and $\kappa=\tilde{\kappa}-{T \alpha^2}/{\sigma}$ is the thermal conductivity measured for vanishing particle current, ${\bf j}=0$. We see that under the condition $T(\zeta\sigma-\chi\alpha)^2/(c_n\chi^2\sigma)\ll (\sigma/\chi,\kappa/c_n)$ the denominator in Eq.~\eqref{eq:factor} approximately factorizes. Both $\zeta$ and $\alpha$ vanish for particle hole-symmetric systems. As a consequence, the left-hand side of the mentioned inequality can be estimated to be smaller by a factor of $(T/\mu)^2$ compared to the right hand side. Within the accuracy of our calculation, we may therefore neglect the last term in the expression for $\tr[\hat{D}]$ in Eq.~\eqref{eq:dettr} \cite{ckappa_Remark}. It is then natural to introduce the diffusion coefficients for density and heat as $D_n=\sigma/\chi$ and $D_k=\kappa/c_n$, respectively. In this way, the density-density correlation function is obtained in the form $\chi_{nn}(\bfq,\omega)=-\chi D_n\bfq^2/(D_n\bfq^2-i\omega)$, and the heat density-density correlation function takes the form given in Eq.~\eqref{eq:chikndiff} of the main text. 

\section{Comment on the connection to heat density-heat density correlation function}
In this appendix, we explain how the heat density-density correlation function is connected to heat density-heat density correlation function. 
\subsection{Connection to heat vertex correction}
\label{sec:connHeat}
We would like to mention a relation between the heat vertex corrections for the heat density-density correlation function, $\delta \chi_{kn,\mathcal{H}}$, and those for the heat density-heat density correlation function, $\delta \chi_{kk,\mathcal{H}}$ (this notation refers to the vertex correction for \emph{one} vertex only). For the corrections for which $D_\eps'$ is extracted from the slow diffuson in the calculation of $\delta \chi_{kn,\mathcal{H}}$, i.e. the diffuson carrying the external frequency $\omega$ and momentum $\bfq$, the following relation exists
\begin{align}
&\left.\delta \chi_{kn,\mathcal{H}}(\bfq,\omega)\right|_{\mathcal{D}^\eps_{\bfq,\omega}\rightarrow \mD_{\bfq,\omg} - \eps D_\eps' \bfq^2 \mD^{2}_{\bfq,\omg}}\no\\
&=- \mD_{\bfq,\omg} D_\eps' \bfq^2 \delta \chi_{kk,\mathcal{H}}(\bfq,\omega).
\end{align}
The existence of this relationship between the two correlation functions provides a useful check for the calculation of $\chi_{k n}$. A similar relation also exists for the corrections to the diffuson, as will be discussed below.

\subsection{Connection to diffuson correction}
\label{sec:connDiff}
The heat density-heat density correlation function and certain corrections to the heat density-density correlation function are also connected. This relation can be expressed as follows
\begin{align}
&\left.\delta \chi_{kn,\mathcal{D}}(\bfq,\omega)\right|_{\mathcal{D}^\eps_{\bfq,\omega}\rightarrow \mD_{\bfq,\omg} - \eps D_\eps' \bfq^2 \mD^{2}_{\bfq,\omg}}\no\\
&=- 2 \mD_{\bfq,\omg} D_\eps' \bfq^2 \delta \chi_{kk,\mathcal{D}}^{\op{dyn}}(\bfq,\omega). \label{diffuson_kntokkRelation}
\end{align}
Here, the factor of 2 arises from the fact that two slow diffusons are involved in the calculation of the correction to diffuson. We emphasize that relation \eqref{diffuson_kntokkRelation} holds for each diagram, but only for the contribution for which $D_\eps'$ is extracted from the slow diffuson in the calculation of $\delta \chi_{kn,\mathcal{D}}$.

\section{Comment on Hikami box diagrams} 
\label{sec:hikami}
\begin{figure}[b]
\includegraphics[width=8.5cm]{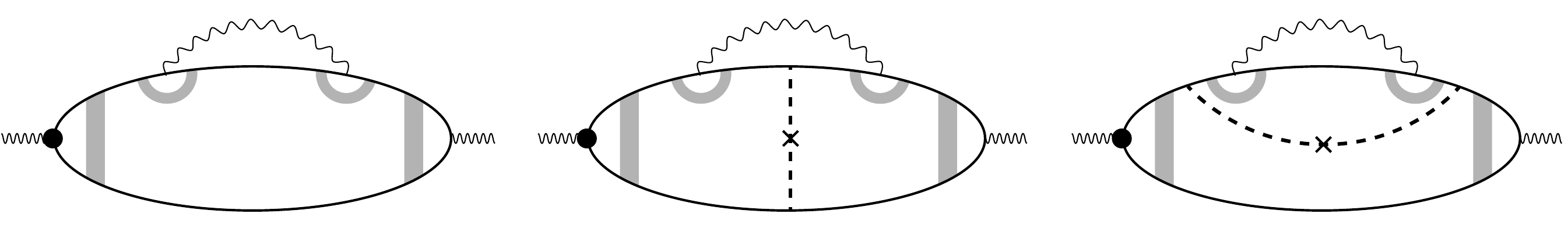}
\caption{Hikami box diagrams contributing to $\delta \chi_{kn,\mathcal{D}}$ in conventional diagrammatic perturbation theory.}
\label{fig:diagrams_hikami_conv}
\end{figure}

One of the diagrams contributing to the corrections to the diffuson, the diagram displayed in Fig.~1(a) of paper I, contains a Hikami box. The Hikami box is a diagrammatic block that describes the coupling of four diffusons. Since the corresponding correction, $\delta \chi_{kn,\mathcal{D}_a}$, plays an important role for the calculation of interaction corrections to $L$, we would like to provide some details concerning the relation between the Hikami box diagrams in the NL$\sigma$M formalism and in the conventional many-body formalism for disordered electron systems. In Fig.~\ref{fig:diagrams_hikami_conv}, we display the Hikami box diagrams for $\delta \chi_{kn}$ as they appear in conventional many-body perturbation theory after applying the so-called cross-technique \cite{abrikosov1975} for implementing the disorder average. In the Keldysh technique, these diagrams correspond to the following analytical expression, which fully accounts for the particle-hole asymmetry
\begin{widetext}
\begin{align}
\delta \chi_{kn}^{\op{Hikami}}(\bfq,\omg) &= \frac{\pi \nu_0 s}{2} \int_{\eps,\bfk,\nu} \eps \Delta_{\eps,\omg} U^R_{\bfk,\nu} (\mD_{\bfq,\omg}^\eps)^2 \no
\\
&\times \left[F_{\eps - \frac{\omg}{2} + \nu} \left(\mD_{\bfk,\nu}^{\eps - \frac{\omg - \nu}{2}}\right)^2 \left(D_{\eps - \frac{\nu}{2}} \bfq^2 + D_{\eps + \frac{\nu}{2} -\omg} \bfk^2 - i (\nu + \omg) - \frac{3}{2} i D_\eps' \bfk^2 D \bfq^2\right)\right. \no
\\
&- \left.F_{\eps + \frac{\omg}{2} - \nu} \left(\mD_{\bfk,\nu}^{\eps + \frac{\omg - \nu}{2}}\right)^2 \left(D_{\eps + \frac{\nu}{2}} \bfq^2 + D_{\eps - \frac{\nu}{2} + \omg} \bfk^2 - i (\nu + \omg) + \frac{3}{2} i D_\eps' \bfk^2 D \bfq^2\right)\right]. \label{eq:hikami}
\end{align}
\end{widetext}
This result was derived in the imaginary time technique in Ref.~\onlinecite{Fabrizio91}. We will now point out how the same result can be obtained in the NL$\sigma$M formalism. We would like to stress, in particular, that the contribution from $\delta \chi^>_{k n, \mathcal{D}_a}(\bfq,\omg)$ is not sufficient to fully account for $\delta \chi_{kn}^{\op{Hikami}}(\bfq,\omg)$ in Eq.~\eqref{eq:hikami}. For the sake of definiteness, we work with the $\zeta=-1$ parameterization, which coincides with the square-root even parameterization in the context of the one-loop calculation. In this parameterization, additional contributions to $\delta \chi_{kn}^{\op{Hikami}}(\bfq,\omg)$  come from $\delta \chi^>_{k n, \mathcal{D}_b}(\bfq,\omg)$, the charge vertex correction $\delta \chi_{k n, \mathcal{C}^>_c}^{\op{dyn}}(\bfq,\omg)$ and the heat vertex correction $\delta \chi_{k n, \mathcal{H}^>_c}^{\op{dyn}}(\bfq,\omg)$. Specifically, these additional terms arise through the substitution $\delta \hat{X} \rightarrow (1/4 i) D_\eps' (\nabla \hat{Q})^2$ in the calculation of Fig.~1(b), Fig.~\ref{fig:diagrams_charge}(c) and Fig.~\ref{fig:diagrams_heat}(c). After adding all these terms, one recovers Eq.~\eqref{eq:hikami}.

\section{Evaluation of integrals}
\label{sec:int}
In this section, we first state and then evaluate all the logarithmic integrals relevant for our calculation. A list of final results can be found at the end of the section.

We first list the logarithmic integrals for which at least one of the two energies $|\nu|$ and $D_0\bfk^2$ lies in the RG energy interval $(T,1/\tau)$. These are the integrals of type $1$ in the classification presented in the introduction to Sec.~\ref{sec:intcorr}. Most of these integrals come in pairs, $I_i$ and $I_i'$. Each pair arises from the same parent integral $\mathcal{I}_i(\eps)$ as 
\begin{align}
\mathcal{I}^\xi_i(\eps)=I_i(\eps)+\xi I_i'(\eps),
\end{align}
where $I_i'$ is proportional to the parameter $D_\eps'$ encoding the particle-hole asymmetry in our model. The integrals $\mathcal{I}^\xi_i(\eps)$ read as 
\begin{align}
\mI_1^{\xi}(\eps) &=- \frac{i}{6} \int_{\bfk,\nu} (F_{\eps+\nu} - F_{\eps-\nu}) (\mD_{\bfk,\nu}^{\xi})^2 U^R_{\bfk,\nu} \label{defInteps}
\\
\mI_{D}^{\xi}(\eps) &= - \frac{i D_{\xi}}{D_0} \int_{\bfk,\nu} (F_{\eps+\nu} - F_{\eps-\nu}) (\mD^{\xi}_{\bfk,\nu})^3 D_{\xi} \bfk^2 U^R_{\bfk,\nu} \no
\\
\mI^{\xi}_z(\eps) &= \frac{1}{2} \int_{\bfk,\nu} \partial_{\eps}(F_{\eps+\nu} + F_{\eps-\nu}) \mD^{\xi}_{\bfk,\nu} \Re U^R_{\bfk,\nu} \no
\\
\mI^{\xi}_2(\eps) &=- \frac{1}{2 \eps} \int_{\bfk,\nu} \nu \partial_{\eps}(F_{\eps+\nu} - F_{\eps-\nu}) \mD^{\xi}_{\bfk,\nu} \Re U^R_{\bfk,\nu} \no
\\
\mI^{\xi}_5(\eps) &= \frac{1}{2 \eps} \int_{\bfk,\nu} (F_{\eps+\nu} + F_{\eps-\nu}) \mD^{\xi}_{\bfk,\nu} \Re U^R_{\bfk,\nu} \no
\\
\mI^{\xi}_6(\eps) &= \frac{1}{2}  \int_{\bfk,\nu}  \nu^2  \partial_{\eps}(F_{\eps+\nu} + F_{\eps-\nu})  \mD_{\bfk,\nu}^\xi  \Re U^R_{\bfk,\nu}.\no
\end{align}
In the calculation of interaction corrections, all of these integrals appear in combination with the window function $\Delta_{\eps,\omg}$, so that the frequency $|\eps|$ can be assumed to be smaller than or of the order of the temperature $T$.

Apart from all the above integrals, we have two separate integrals, $I_3'$ and $I_4'$, which also give corrections from RG energy interval. We start with
\begin{align}
I_{3}'(\eps) &= \frac{1}{4 \eps} \int_{\bfk,\nu} (F_{\eps+\nu} + F_{\eps-\nu}) \mD_{\bfk,\nu}^2 D_\eps' \bfk^2 U^{R}_{\bfk,\nu} \label{eq:I3_def}.
\end{align}
The $I_3'$ integral does not have a partner integral $I_3$ accompanying it like the integrals in Eq.~\eqref{defInteps}. The same is true for the $I_4'$ integral. In fact, the integral $I_4'$ will require a separate consideration due to the its complicated structure
\begin{align}
I_4' &= \frac{i}{c_0 T \omg} \frac{\pi \nu_0 s}{2} \int_{\eps,\bfk,\nu} \Delta_{\eps,\omg} (F_{\eps+\nu} - F_{\eps-\nu}) \nu^2 \mD_{\bfk,\nu}^3 D_\eps' \bfk^2 U^R_{\bfk,\nu}\no
\\
&+ \frac{1}{c_0 T \omg} \frac{\pi \nu_0 s}{4} \int_{\eps,\bfk,\nu} \Delta_{\eps,\omg} (F_{\eps+\nu} - F_{\eps-\nu}) \nu \mD_{\bfk,\nu}^2 D_\eps' \bfk^2 U^R_{\bfk,\nu} \no
\\
&+ \frac{1}{c_0 T \omg} \frac{\pi \nu_0 s}{8} \int_{\eps,\bfk,\nu} \Delta_{\eps,\omg} (F_{\eps+\nu} - F_{\eps-\nu}) \nu \mD^2_{\bfk,\nu} D_\eps' \bfk^2 \times\no\\
&\times(U^R_{\bfk,\nu} + U^A_{\bfk,\nu}) \no
\\
&+ \frac{1}{c_0 T} \frac{\nu_0 s}{4} \int_{\bfk,\nu} B_{\nu} \nu \mD^2_{\bfk,\nu} D_\eps' \bfk^2 (U^R_{\bfk,\nu} -U^A_{\bfk,\nu}) \no
\\
&+ \frac{1}{c_0 T} \frac{\nu_0^2 s^2}{2} \int_{\bfk,\nu} B_{\nu} \nu^3 \mD^4_{\bfk,\nu} D_\eps' \bfk^2 U^{R}_{\bfk,\nu} U^R_{\bfk, \nu} \no
\\
&- \frac{i}{c_0 T} \frac{\nu_0^2 s^2}{2} \int_{\bfk,\nu} B_{\nu} \nu^2 \mD^3_{\bfk,\nu} D_\eps' \bfk^2 U^{R}_{\bfk,\nu} U^{R}_{\bfk,\nu}. 
\end{align}

Now, we proceed to the evaluation of the logarithmic integrals. For the $I_1$ integral, important momenta lie in the interval {$\nu^2/(D \kappa_s^2) < D \bfk^2 < |\nu|$.} After the straightforward momentum integration, the integral can be evaluated as
\begin{align}
I_1(\eps) &\approx \frac{\pi \rho}{12} \int_{\nu} \frac{F_{\eps+\nu} - F_{\eps-\nu}}{\nu} \ln\frac{D_0 \kappa_s^2}{|\nu|}\\
 &\approx \frac{\rho}{6} \ln\frac{1}{\max(|\eps|,T) \tau} \ln\frac{D_0 \kappa_s^2}{\max(|\eps|,T)},\nonumber
\end{align}
where we have only kept the most singular terms. This integral is associated with the correction to the tunneling density of states \cite{AltLee80} and eventually drops out from the overall result. 
 
Further integrals can be evaluated as
{
\begin{align}
I_D(\eps) &\approx \frac{\pi \rho}{2} \int_{\nu} \frac{F_{\eps+\nu} - F_{\eps-\nu}}{\nu} \\
&\approx \rho \ln\frac{1}{\max(|\eps|,T) \tau} \equiv I(\eps), \nonumber
\\
I_D'(\eps) &\approx \frac{\pi \rho}{4 \mu} \int_{\nu} \frac{F_{\eps+\nu} - F_{\eps-\nu}}{\nu} \approx \frac{D_\eps'}{2 D}  I(\eps), \no
\\
I_1'(\eps) &\approx -\frac{\pi \rho}{6 \mu} \int_{\nu} \frac{F_{\eps+\nu} - F_{\eps-\nu}}{\nu} \approx - \frac{D_\eps'}{3 D}  I(\eps),\no
\\
I_z(\eps) &\approx \frac{\pi \rho}{2} \int_{\nu} \partial_\eps F_{\eps+\nu} \ln\frac{1}{|\nu| \tau} \approx \frac{1}{2} I(\eps), \no
\\
I_z'(\eps) &\approx -\frac{\pi \rho}{2 \mu} \int_{\nu} \partial_\eps F_{\eps+\nu} \ln\frac{1}{|\nu| \tau} \approx -\frac{D_\eps'}{2 D} I(\eps). \no
\end{align} 

Here, for the $I_D, I_D'$ and $I_1'$ integrals, relevant $D \bfk^2$ and $|\nu|$ lie within the RG interval $(T,1/\tau)$. By contrast, in the $I_z$ and $I_z'$ integrals, relevant $D \bfk^2$} lie in the RG energy interval, while $|\nu|$ are smaller or of the order of the temperature. The same is true for the integrals $I_2(\eps),I_2'(\eps), I_{3}'(\eps), I_5(\eps)$ and $I_5'(\eps)$, for which we just state the results
{
\begin{align}
I_2(\eps) &\approx \frac{1}{2} I(\eps), \\
I_2'(\eps) &\approx -\frac{D_\eps'}{2 D} I(\eps),\nonumber\\
I_3'(\eps) &\approx \frac{D_\eps'}{4 D} I(\eps), \nonumber\\
I_5(\eps) &\approx \frac{1}{2} I(\eps),\nonumber\\
I_5'(\eps) &\approx -\frac{D_\eps'}{2 D} I(\eps). \nonumber
\end{align}
}
Now, we will evaluate the integral $I_4'$. The $\eps$ integrals can be performed with the help of the identity $\int_\eps \Delta_{\eps,\omg} (F_{\eps+\nu} - F_{\eps-\nu}) = \frac{2 \omg}{\pi} B_{\nu} + \frac{2 \omg}{\pi} B'_{\nu} \nu$. In this way, we arrive at the following expression
\begin{align}
I_4' &= \frac{i \nu_0 s}{c_0 T} \int_{\bfk,\nu} (\partial_\nu B_{\nu}) \nu^3 \mD^3_{\bfk,\nu} D_\eps' \bfk^2 U^R_{\bfk,\nu} 
\\
&+ \frac{\nu_0 s}{2 c_0 T} \int_{\bfk,\nu} (\partial_\nu B_{\nu}) \nu^2 \mD^2_{\bfk,\nu} D_\eps' \bfk^2 U^R_{\bfk,\nu} \no
\\
&+ \frac{\nu_0 s}{4 c_0 T} \int_{\bfk,\nu} (\partial_\nu B_{\nu}) \nu^2 \mD^2_{\bfk,\nu} D_\eps' \bfk^2 (U^R_{\bfk,\nu} + U^A_{\bfk,\nu}) \no
\\
&+ \frac{\nu_0 s}{c_0 T} \int_{\bfk,\nu} B_{\nu} \nu \mD^2_{\bfk,\nu} D_\eps' \bfk^2 U^R_{\bfk,\nu} \no
\\
&+ \frac{i \nu_0 s}{c_0 T} \int_{\bfk,\nu} B_{\nu} \nu^2 \mD^3_{\bfk,\nu} D_\eps' \bfk^2 U^R_{\bfk,\nu} \no
\\
&- \frac{i \nu_0^2 s^2}{2 c_0 T} \int_{\bfk,\nu} B_{\nu} \nu^2 \mD^4_{\bfk,\nu} D_\eps' D \bfk^4 U^{R}_{\bfk,\nu} U^{R}_{\bfk,\nu}. \no
\end{align}

For the first three terms involving the derivative of the bosonic distribution function, $\partial_\nu B_\nu$, one can perform the $\bfk$ and $\nu$ integrals in the same way as for the $I_z'$ integral. Taken together, these three terms give $- 2 \partial_\mu z$. On the other hand, the last three terms containing the factor $B_\nu$ have been considered previously in Ref.~\onlinecite{Schwiete21}, and are equal to $- (1/c_0 T) (- \partial_\mu k^{\op{dm}} - n^{\op{dm}}) =- (1/c_0 T) \delta\chi_{kn}^{\op{st}}$, so that we can conclude
{
\begin{align}
I_4' &\approx- \frac{D_\eps'}{2 D} \rho \ln\frac{1}{T \tau} =- \partial_\mu z.
\end{align}
}
For the last two integrals, $I_6(\eps)$ and $I_6'(\eps)$, it is convenient to evaluate these two integrals by performing the $\eps$ integral first, i.e., to evaluate $\int_\eps \Delta_{\eps,\omg} I_6(\eps)$ and $\int_\eps \Delta_{\eps,\omg} I_6'(\eps)$. After performing the $\eps$ integral with the help of the identity $B_{\eps-\eps'} (F_{\eps} - F_{\eps'}) = 1 - F_{\eps} F_{\eps'}$, we can evaluate the $\bfk$ and the $\nu$ integral similar to $I_4'$ integral.
{
\begin{align}
I_6 &= \rho \ln \frac{1}{T \tau},  \quad I_6' =- \frac{D_\eps'}{D} \rho \ln \frac{1}{T \tau}.
\end{align}
}

The next set of integrals gives rise to logarithmic corrections from the sub-thermal energy interval. These are the integrals of type $2$ in the classification presented in the introduction to Sec.~\ref{sec:intcorr}. 
\begin{align}
I^h_1(\eps) &=- \frac{i}{\eps} \int_{\bfk,\nu} \nu (F_{\eps+\nu} + F_{\eps-\nu}) \mD_{\bfk,\nu}^2 U^R_{\bfk,\nu}, 
\\
\tilde{I}^h_1(\eps) &= \frac{1}{\eps} \int_{\bfk,\nu} \nu (F_{\eps+\nu} + F_{\eps-\nu}) \mD_{\bfk,\nu}^2 \Im U^R_{\bfk,\nu}, \no
\\
I^h_2(\eps) &= \frac{i}{2 \eps} \int_{\bfk,\nu} (F_{\eps+\nu} + F_{\eps-\nu}) \mD_{\bfk,\nu} U^{R}_{\bfk,\nu}, \no
\\
I^h_3(\eps) &=- \frac{1}{\eps} \int_{\bfk,\nu} \nu (F_{\eps+\nu} + F_{\eps-\nu}) \mD_{\bfk,\nu} \bar{\mD}_{\bfk,\nu} U^{R}_{\bfk,\nu}, \no
\\
I^h_{4}(\eps) &= \frac{i}{\eps} \int_{\bfk,\nu} \nu (F_{\eps+\nu} + F_{\eps-\nu}) \mD_{\bfk,\nu}^2 U^A_{\bfk,\nu}, \no
\\
I^h_{5}(\eps) &= 2  \int_{\bfk,\nu}  \nu^2  (2 B_{\nu} - (F_{\eps+\nu} - F_{\eps-\nu}))  \mD_{\bfk,\nu}^2 \Im U^{R}_{\bfk,\nu}.\no
\end{align}
All the $\nu$ integrals above are constrained by the presence of the factor $(F_{\eps+\nu} + F_{\eps-\nu})$, which restricts relevant $|\nu|$ to values smaller than or of the order of the temperature. {Important $D \bfk^2$ for these integrals lie in the range $\nu^2/(D \kappa_s^2) < D \bfk^2 < |\nu|$.  Let us showcase one such integral
\begin{align}
&\tilde{I}^h_1(\eps) =- \frac{i}{2 \eps} \int_{\bfk,\nu} \nu (F_{\eps+\nu} + F_{\eps-\nu}) \mD_{\bfk,\nu}^2 (U^R_{\bfk,\nu} - U^A_{\bfk,\nu})\nonumber\\
& \approx \frac{\pi \rho}{2 \eps} \int_{\nu} (F_{\eps+\nu} + F_{\eps-\nu}) \ln\frac{D \kappa_s^2}{|\nu|} \approx \rho \ln\frac{D \kappa_s^2}{\max(|\eps|,T)}
\end{align}
}
The final results for the other integrals (except for $I_5^h$) read as
{
\begin{align}
I^h_1(\eps) &\approx \rho \ln\frac{D \kappa_s^2}{\max(|\eps|,T)} = I^h(\eps), \\
I^h_{2}(\eps) &\approx \frac{1}{2} I^h(\eps),\nonumber\\
I^h_3(\eps) &\approx I^h(\eps), \\
I^h_{4}(\eps) &\approx I^h(\eps).
\end{align}
}
The integral $I_5^h$ is treated in a similar way as the integral $I_6$. So, we first perform the $\eps$ integral ($\int_\eps \Delta_{\eps,\omg} I_5^h(\eps)$) and then evaluate the $\bfk$ and the $\nu$ integrals
{
\begin{align}
I_5^h = 4 \rho \ln \frac{D \kappa_s^2}{T}.
\end{align}
}
\paragraph*{List of results:}
We will now present a list of results for all the relevant integrals. We note that the integrals denoted as $I_i(\eps)$, $I_i'(\eps)$, $\tilde{I}^h_1(\eps)$ and $I^h_i(\eps)$ are weakly frequency-dependent. However as discussed before, for the purpose of our calculation it is sufficient to evaluate these integrals in the limit $\eps \rightarrow 0$. Therefore, we will use the notation: $I_i = I_i(\eps \rightarrow 0)$.  

We first list the integrals $I_1$ and $I_1'$, related to the tunneling density of states, and $I_D$ and $I_D'$, related to the diffusion coefficient
{
\begin{align}
I_1 &= \frac{1}{6} \rho \ln \frac{1}{T \tau} \ln \frac{D \kappa_s^2}{T} ,  \quad I_1' =- \frac{D_\eps'}{3 D} \rho \ln \frac{1}{T \tau}, \label{eq:I1}
\\
I_D &= \rho \ln \frac{1}{T \tau} =- \frac{\delta D}{D},  \quad I_D' = \frac{D_\eps'}{2 D} \rho \ln \frac{1}{T \tau}. \label{eq:ID}
\end{align}
}
The integrals associated with the frequency renormalization $z$ are
{
\begin{align}
I_z &= \frac{1}{2} \rho \ln \frac{1}{T \tau}, \quad I_z' =- \frac{D_\eps'}{2 D} \rho \ln \frac{1}{T \tau}, \label{eq:Iz}
\\
I_2 &= I_5 = \frac{I_6}{2} = I_z = - \delta z, \no
\\
I_2' &= I_4' = I_5' =- 2 I_{3}' = \frac{I_6'}{2} = I_z' = - \partial_\mu z. \no
\end{align}
}
We have already addressed all the integrals with main contributions from the RG energy interval. Now, we list the corrections from the sub-thermal energy interval
{
\begin{align}
I_1^h &= \tilde{I}_1^h = I_{3}^h = I_{4}^h = 2 I_2^h = \frac{I_5^h}{4} =I^h = \rho \ln \frac{D \kappa_s^2}{T}. \label{eq:Ih}
\end{align}
}

\section{Cancellation of $J$ terms}
\label{sec:Jterms}

In this section, we demonstrate the cancellation of the $J$ terms, which were introduced at the beginning of Sec.~\ref{sec:intcorr}. These terms have a structure that is inconsistent with the general form of the heat density-density correlation function in Eq.~\eqref{eq:chikndiff} and Eq.~\eqref{eq:strucPert}. Indeed, their presence in the overall result would signal the formation of a gap in the diffuson, 
\begin{align}
\delta \chi_{kn}^{m}(\bfq,\omg) &\propto \int_{\eps} \frac{\eps \Delta_{\eps,\omg}}{D(\eps) \bfq^2 - i \omg +\Gamma(\eps)}\nonumber\\
 &\approx 2 i \omg D_\eps' \bfq^2 \mD^3_{\bfq,\omg} \Gamma_0 -  i \omg \mD^2_{\bfq,\omg} \Gamma_\eps' + ....\label{eq:mass}
\end{align}
and a violation of the conservation laws for charge and energy. Here, {$D(\eps) = D + \eps D_\eps'$} and $\Gamma(\eps) = \Gamma_0 + \eps \Gamma_\eps'$ are the diffusion coefficient and the frequency dependent gap, respectively, in the presence of the particle-hole asymmetry. Terms of the form displayed on the right hand side of Eq.~\eqref{eq:mass} can arise for individual diagrams, but they must cancel out from the overall calculation. Here, we present details of this cancellation. Similar terms also arise in intermediate steps of the calculation for the density-density and heat density-heat density correlation function, as discussed in detail in Ref.~\onlinecite{Schwiete16a}. 

Now we list contributions to the $J$ terms classified according to their origin. We use the convention that $\delta \chi_{k n, \mD_a}^{J} (\bfq, \omega)$ represents $J$ terms originating from Fig.~1(a).

\begin{widetext}
\begin{align}
\delta\chi_{k n,\mD_a}^{J}(\bfq,\omg) &= \zeta \pi \nu_0 s \mD_{\bfq,\omega}^3 D_{\eps}' \bfq^2 \int_{\eps,\bfk,\nu} \eps^2 \Delta_{\eps,\omg} (F_{\eps+\nu} - F_{\eps-\nu}) \mD_{\bfk,\nu} U^R_{\bfk,\nu} 
\\
&+ \frac{\zeta \pi \nu_0 s}{2} \mD_{\bfq,\omg}^2 \int_{\eps,\bfk,\nu} \eps^2 \Delta_{\eps,\omg} (F_{\eps+\nu} - F_{\eps-\nu}) \mD_{\bfk,\nu}^2 D_{\eps}' \bfk^2 U^{R}_{\bfk,\nu}\no
\\
&+ \frac{\zeta \pi \nu_0 s}{4} \mD_{\bfq,\omg}^2 \int_{\eps,\bfk,\nu} \eps \nu \Delta_{\eps,\omg} (F_{\eps+\nu} + F_{\eps-\nu}) \mD_{\bfk,\nu}^2 D_{\eps}' \bfk^2 U^{R}_{\bfk,\nu}, \no
\end{align}

\begin{align}
\delta\chi_{k n,\mD_b}^{J}(\bfq,\omg) &= - (1 + \zeta) \pi \nu_0 s \mD_{\bfq,\omega}^{3} D_{\eps}' \bfq^2 \int_{\eps,\bfk,\nu} \eps^2 \Delta_{\eps,\omg} (F_{\eps+\nu} - F_{\eps-\nu}) \mD_{\bfk,\nu} U_{\bfk,\nu}^{R} 
\\
&- \frac{(1 + \zeta) \pi \nu_0 s}{2} \mD_{\bfq,\omega}^{2} \int_{\eps,\bfk,\nu} \eps^2 \Delta_{\eps,\omg} (F_{\eps+\nu} - F_{\eps-\nu}) \mD_{\bfk,\nu}^2 D_{\eps}' \bfk^2 U_{\bfk,\nu}^{R}\no
\\
&- \frac{(1 + \zeta) \pi \nu_0 s}{4} \mD_{\bfq,\omega}^{2} \int_{\eps,\bfk,\nu} \eps \nu \Delta_{\eps,\omg} (F_{\eps+\nu} + F_{\eps-\nu}) \mD_{\bfk,\nu}^2 D_{\eps}' \bfk^{2} U_{\bfk,\nu}^{R}. \no
\end{align}
We see that combining the contributions from Fig.~1(a) and (b) leads to the cancellation of the terms proportional to the parameter $\zeta$, so that
\begin{align}
\delta\chi_{k n,\mD_{a-b}}^{J}(\bfq,\omg) &= - \pi \nu_0 s \mD_{\bfq,\omega}^{3} D_{\eps}' \bfq^2 \int_{\eps,\bfk,\nu} \eps^2 \Delta_{\eps,\omg} (F_{\eps+\nu} - F_{\eps-\nu}) \mD_{\bfk,\nu} U_{\bfk,\nu}^{R} 
\\
&- \frac{\pi \nu_0 s}{2} \mD_{\bfq,\omega}^{2} \int_{\eps,\bfk,\nu} \eps^2 \Delta_{\eps,\omg} (F_{\eps+\nu} - F_{\eps-\nu}) \mD_{\bfk,\nu}^2 D_{\eps}' \bfk^2 U_{\bfk,\nu}^{R}\no
\\
&- \frac{\pi \nu_0 s}{4} \mD_{\bfq,\omega}^{2} \int_{\eps,\bfk,\nu} \eps \nu \Delta_{\eps,\omg} (F_{\eps+\nu} + F_{\eps-\nu}) \mD_{\bfk,\nu}^2 D_{\eps}' \bfk^{2} U_{\bfk,\nu}^{R}. \no
\end{align}
Similar cancellations occur for the $J$ terms corresponding to Fig.~1(c) and (d). This is why we group them together
\begin{align}
\delta\chi_{k n,\mD_{c-d}}^{J}(\bfq,\omg) &= \pi \nu_0 s \mD_{\bfq,\omg}^{3} D_{\eps}' \bfq^2 \int_{\eps,\bfk,\nu} \eps^2 \Delta_{\eps,\omg} (F_{\eps+\nu} - F_{\eps-\nu}) \mD_{\bfk,\nu} U^{R}_{\bfk,\nu} 
\\
&+ 3 i \pi \nu_0 s \mD_{\bfq,\omg}^3 D_{\eps}' \bfq^2 \int_{\eps,\bfk,\nu} \eps \nu \Delta_{\eps,\omg} (F_{\eps+\nu} + F_{\eps-\nu}) \mD_{\bfk,\nu} \Im U^{R}_{\bfk,\nu} \no
\\
&- i \pi \nu_0 s \mD_{\bfq,\omg}^3 D_{\eps}' \bfq^2 \int_{\eps,\bfk,\nu} \nu^2 \Delta_{\eps,\omg} \left(2 B_{\nu} - (F_{\eps+\nu} - F_{\eps-\nu}) \right) \mD_{\bfk,\nu} \Im U^{R}_{\bfk,\nu} \no
\\
&+ \frac{\pi \nu_0 s}{2} \mD_{\bfq,\omg}^{2} \int_{\eps,\bfk,\nu} \eps^2 \Delta_{\eps,\omg} (F_{\eps+\nu} - F_{\eps-\nu}) \mD_{\bfk,\nu}^2 D_{\eps}' \bfk^2 U^{R}_{\bfk,\nu}\no
\\
&+ \frac{\pi \nu_0 s}{4} \mD_{\bfq,\omg}^{2} \int_{\eps,\bfk,\nu} \eps \nu \Delta_{\eps,\omg} (F_{\eps+\nu} + F_{\eps-\nu}) \mD_{\bfk,\nu}^2 D_{\eps}' \bfk^2 \left(3 U^{R}_{\bfk,\nu} - 2 U^{A}_{\bfk,\nu}\right) \no
\\
&- \frac{i \pi \nu_0 s}{2} \mD_{\bfq,\omg}^2 \int_{\eps,\bfk,\nu} \nu^2 \Delta_{\eps,\omg} \left(2 B_{\nu} - (F_{\eps+\nu} - F_{\eps-\nu}) \right) \mD_{\bfk,\nu}^2 D_{\eps}' \bfk^2 \Im U^{R}_{\bfk,\nu}. \no
\end{align}
We can add all the contributions from Fig.~1(a), (b), (c) and (d) to arrive at the following expression
\begin{align}
\delta\chi_{k n,\mD_{a-d}}^{J}(\bfq,\omg) &= 3 i \pi \nu_0 s \mD_{\bfq,\omg}^3 D_{\eps}' \bfq^2 \int_{\eps,\bfk,\nu} \eps \nu \Delta_{\eps,\omg} (F_{\eps+\nu} + F_{\eps-\nu}) \mD_{\bfk,\nu} \Im U^{R}_{\bfk,\nu} 
\\
&- i \pi \nu_0 s \mD_{\bfq,\omg}^3 D_{\eps}' \bfq^2 \int_{\eps,\bfk,\nu} \nu^2 \Delta_{\eps,\omg} \left(2 B_{\nu} - (F_{\eps+\nu} - F_{\eps-\nu}) \right) \mD_{\bfk,\nu} \Im U^{R}_{\bfk,\nu} \no
\\
&+ i \pi \nu_0 s \mD_{\bfq,\omg}^{2} \int_{\eps,\bfk,\nu} \eps \nu \Delta_{\eps,\omg} (F_{\eps+\nu} + F_{\eps-\nu}) \mD_{\bfk,\nu}^2 D_{\eps}' \bfk^2 U^{R}_{\bfk,\nu}\no
\\
&- \frac{i \pi \nu_0 s}{2} \mD_{\bfq,\omg}^2 \int_{\eps,\bfk,\nu} \nu^2 \Delta_{\eps,\omg} \left(2 B_{\nu} - (F_{\eps+\nu} - F_{\eps-\nu}) \right) \mD_{\bfk,\nu}^2 D_{\eps}' \bfk^2 \Im U^{R}_{\bfk,\nu}. \no
\end{align}
The final contribution comes from the drag-like diagrams in Fig.~1(e-h).
\begin{align}
\delta\chi_{k n,\mD_{e-h}}^{J}(\bfq,\omg) &= - i \pi \nu_0 s \mD^{3}_{\bfq,\omg} D_{\eps}' \bfq^2 \int_{\eps,\bfk,\nu} \eps \nu \Delta_{\eps,\omg} (F_{\eps+\nu} + F_{\eps-\nu}) \mD_{\bfk,\nu} \Im U^{R}_{\bfk,\nu} 
\\
&- i\pi \nu_0 s \mD^2_{\bfq,\omg} \int_{\eps,\bfk,\nu} \eps \nu \Delta_{\eps,\omg} (F_{\eps+\nu} + F_{\eps-\nu}) \mD^2_{\bfk,\nu} D_{\eps}' \bfk^2 \Im U^{R}_{\bfk,\nu}\no
\\
&+ \frac{i \pi \nu_0 s}{2} \mD^2_{\bfq,\omg} \int_{\eps,\bfk,\nu} \nu^2 \Delta_{\eps,\omg} \left(2 B_{\nu} - (F_{\eps+\nu} - F_{\eps-\nu}) \right) \mD^2_{\bfk,\nu} D_{\eps}' \bfk^2 \Im U^{R}_{\bfk,\nu}. \no
\end{align}
Adding all the contributions arising from the calculation of the corrections to the diffuson (see Fig.~1), we obtain
\begin{align}
\chi_{k n,\mD}^{J}(\bfq,\omg) &= 2 i \pi \nu_0 s \mD^{3}_{\bfq,\omg} D_{\eps}' \bfq^2 \int_{\eps,\bfk,\nu} \eps \nu \Delta_{\eps,\omg} (F_{\eps+\nu} + F_{\eps-\nu}) \mD_{\bfk,\nu} \Im U^{R}_{\bfk,\nu}
\\
&- i \pi \nu_0 s \mD_{\bfq,\omg}^3 D_{\eps}' \bfq^2 \int_{\eps,\bfk,\nu} \nu^2 \Delta_{\eps,\omg} \left(2 B_{\nu} - (F_{\eps+\nu} - F_{\eps-\nu}) \right) \mD_{\bfk,\nu} \Im U^{R}_{\bfk,\nu}. \no
\end{align}
We can use the identity $B_{\eps-\eps'} (F_{\eps} - F_{\eps'}) = 1 - F_{\eps} F_{\eps'}$ to derive the following relations
\begin{align}
\int_{\eps} \eps \Delta_{\eps,\omg} (F_{\eps+\nu} + F_{\eps-\nu}) &\approx - \frac{\omg}{\pi} B_\nu' \nu^2 + \mathcal{O}(\omg^3), \quad \int_{\eps} \Delta_{\eps,\omg} (2 B_\nu -(F_{\eps+\nu} - F_{\eps-\nu})) \approx - \frac{2 \omg}{\pi} B_\nu' \nu + \mathcal{O}(\omg^3). 
\end{align}
With the help of these relations, we can immediately establish the overall cancellation of the $J$ terms 
\begin{align}
\chi_{k n,\mD}^{J}(\bfq,\omg) = 0.
\end{align}
\end{widetext}


\begin{thebibliography}{58}
\bibitem [{not()}]{notation_remark}%
  \BibitemOpen
  \href@noop {} {}\bibinfo {note} {We set $\hbar = c = k_B = 1$ throughout the
  paper.}\BibitemShut {Stop}%
\bibitem [{Cha()}]{Charge_Remark}%
  \BibitemOpen
  \href@noop {} {}\bibinfo {note} {For the sake of notational simplicity, we
  set $e=1$ in this section.}\BibitemShut {Stop}%
\bibitem [{cka()}]{ckappa_Remark}%
  \BibitemOpen
  \href@noop {} {}\bibinfo {note} {For similar reasons we may neglect the
  differences between $c_n$ and $c_\mu$ and between $\tilde{\kappa}$ and
  $\kappa$.}\BibitemShut {Stop}%
\end{thebibliography}

%

\end{document}